\documentclass[useAMS,usenatbib]{mn2e}

\usepackage{psfig, epsf, epsfig}

\title[Chemical evolution with dust wind]{Chemical evolution of galaxies with
radiation-driven dust wind}
\author[K. Bekki and T. Tsujimoto]
{Kenji Bekki${}^1$\thanks{E-mail:
bekki@cyllene.uwa.edu.au} 
and Takuji Tsujimoto${}^2$ \\
${}^1$ICRAR M468
The University of Western Australia
35 Stirling Hwy, Crawley
Western Australia 6009, Australia \\
${}^2$National Astronomical Observatory of Japan, Mitaka-shi, Tokyo 181-8588, Japan \\}

\begin{document}

\date{Accepted, Received 2005 February 20; in original form }

\pagerange{\pageref{firstpage}--\pageref{lastpage}} \pubyear{2005}

\maketitle

\label{firstpage}

\begin{abstract}
We discuss how the removal of interstellar dust by radiation pressure of stars
influences the chemical evolution of galaxies by using a new one-zone chemical evolution
models with dust wind.
The removal efficiency of an element (e.g., Fe, Mg, and Ca)
through radiation-driven dust wind in a galaxy is assumed to depend both
on the dust depletion level of the element in interstellar medium
and the total luminosity of the galaxy in the
new model.  We particularly focus on the time evolution of [$\alpha$/Fe]
and its dependence on model parameters for dust wind  in this
study.
The principal results are as follows.
The time evolution of [Ca/Fe] is significantly different between models with and without
dust wind in the sense that [Ca/Fe] can be systematically lower in the models with dust wind.
The time evolution of [Mg/Fe], on the other hand, can not be so different between
the models with and without dust wind owing to the lower level of dust depletion
for Mg.  As a result of this, [Mg/Ca] can be systematically higher in the models with dust wind.
We compare  these results with the observed elemental features  of
stars in 
the Large Magellanic Cloud (LMC),
because a growing number of observational studies on [$\alpha$/Fe] for the LMC
have been recently accumulated for a detailed comparison.
Based on the present new results, we also discuss the origins of [$\alpha$/Fe] 
in the  Fornax dwarf galaxy  and elliptical galaxies in the context of radiation-driven dust wind.
\end{abstract}


\section{Introduction}

Interstellar dust is a fundamentally important component
of interstellar medium (ISM) in galaxies,
because it can control interstellar chemistry of variously different elements,
drive the formation of molecular hydrogen, and modify the spectral energy distributions of galaxies
through absorption and emission  of stellar light.
Dust can form from gas and metals ejected from supernovae (SNe) and Asymptotic giant branch
(AGB) stars,
grow through accretion of metals onto pre-existing grains,
and be destroyed by a number of physical processes such as SN explosions  and shocks
(e.g., Jones et al. 1995; See Hirashita 2013 for a recent review on physical processes
related to dust evolution).
Therefore, the time evolution of star formation rates
and physical conditions of ISM can cause 
rather complicated evolution processes of dust in galaxies.
Although the complicated evolution processes of dust in galaxies have been investigated both by
one-zone chemical evolution models (e.g., Dwek 1998) and numerical simulations (e.g., Bekki 2013; Yozin \& Bekki 2014),
they are yet to be fully understood.

Radiation pressure of stars has long been considered to be 
one of key physical processes of dust evolution
in galaxies (e.g., Chiao \& Wickramasinghe 1972; Barsella et al. 1989;
Ferrara et al.1991, F91; Aguirre et al. 2001). 
Radiation pressure of stars on dust grains is demonstrated to cause dust wind in star-forming
disk galaxies
and the dust wind can be responsible both for the formation of the dusty halos around galaxies
and for the origin of intergalactic dust (e.g., F91).
Such efficient removal of dust from galaxies through radiation-driven wind could cause significant
changes of chemical abundance pattern in galaxies owing to the observed different levels
of dust depletion among different chemical elements (e.g., no depletion for N and severe depletion
for Ca). One of obvious effects of dust removal on galaxy evolution is that
the formation efficiency of ${\rm H_2}$ can be significantly lowered  by the reduced amount
of dust in ISM.

Chemical evolution of galaxies should be strongly influenced by the removal of dust through
radiation-driven
dust wind, because a large fraction of metal is observed to be locked up in dust  of ISM
(e.g., Savage \& Sembach 1996, SS96;  Draine 2009; Jenkins 2009).
However, previous one-zone chemical evolution models with dust formation and destruction
did not consider this important effect of radiation-driven dust wind
on chemical evolution (e.g., Dwek 1998;
Hirashita 1999; Calura et al. 2008; Pipino et al. 2010).
Recent chemodynamical models of galaxy formation with dust evolution did not incorporate
the radiation pressure of stars on dust grains in a self-consistent manner 
(Bekki 2013, 2014),
though dust removal through supernova wind is included.
Therefore, it is largely unclear how the dust wind by radiation pressure of stars can influence
the chemical evolution of galaxies.

A growing number of observational studies have revealed some evidences  
for the existence of dust far beyond galactic disks and for
the outflow of dust in actively star-forming galaxies
(e.g., Holwerda et al. 2009;  Roussel et al. 2010; M\'enard et al. 2010;
Yoshida et al. 2011;  Peek et al. 2013).
For example, M\'enard et al. (2010) detected the presence of dust at $R \sim 20$ kpc
to several Mpc by investigating the correlations between the brightness 
of $\sim 85000$ quasars and the positions of $\sim 2.4 \times 10^6$ galaxies.
They also found that (i) the cosmic dust density is
$\Omega_{\rm dust} \sim 5 \times 10^{-6}$
and (ii) roughly the half of the cosmic dust can come from dust in the
halos of luminous ($L \sim L^{\ast }$) galaxies.
The observed large amount of dust in galaxy halos 
strongly suggests that dust removal from the main bodies of galaxies
needs to be seriously considered in theoretical models of dust and metal
evolution in galaxies.

Recently Pomp\'eia et al. 2008 (P08) investigated [$\alpha$/Fe] of stars in
the Large Magellanic Cloud (LMC) and found that [Ca/Fe] is significantly (by $0.2-0.4$ dex)
underabundant in comparison with other [$\alpha$/Fe]
(e.g., [Mg/Fe]). Although Bekki \& Tsujimoto (2012, BT12) have tried to reproduce
the observed very low [Ca/Fe] and almost solar [Mg/Fe] in the LMC by using their
chemical evolution models with canonical IMFs, they have failed to reproduce
such unique chemical abundances of the LMC. They have therefore concluded that
some physical processes which can lead to the selective removal of Ca would
be required for explaining self-consistently both the observed [Mg/Ca] and [Ca/Fe]
of the LMC stars. Furthermore, unusually low [Ca/Fe] has been observed for the stars
in the Fornax dwarf galaxy (e.g., Letarte et al. 2010). If these low [Ca/Fe]
in massive dwarf galaxies are real, then the origin  needs to be clarified
by theoretical models of galaxy formation and evolution.

The purpose of this paper is thus to construct a new chemical evolution model
with radiation-driven dust wind and thereby to discuss how the dust removal from galaxies
can influence galactic  chemical evolution.
We particularly focus on the influences of dust wind on the evolution of [$\alpha$/Fe],
because the observed dust depletion levels of these $\alpha$ elements are quite different
with one another (e.g., SS96). We compare the results of our new models
mainly with the corresponding observations for the LMC in the present
study, firstly because our previous models (BT12) failed to explain
the observed [Ca/Fe]-[Fe/H] and [Mg/Fe]-[Fe/H] relations self-consistently by using
a canonical model without dust wind,
and secondly because a growing number of observations have been accumulated
for the LMC that can allow us to make a detailed comparison between observations and 
models (e.g., Colucci et al. 2012, C12; Haschke et al. 2012;
Van der Swaelmen et al. 2013, V13).

The plan of the paper is as follows: In the next section,
we describe our new one-zone chemical evolution models
with radiation-driven dust wind.
In \S 3, we
present the results
on the [$\alpha$/Fe] evolution and its dependence on model parameters for dust wind.
In this section, we particularly discuss correlations between [Mg/Ca] and [Fe/H]
in  the models with and without dust wind.
In \S 4, we discuss the latest observational results on the [Ca/Fe]-[Fe/H] relations
of the Fornax dwarf galaxy and lower [Ca/Fe] in elliptical galaxies.
We summarize our  conclusions in \S 5.
Although there are numerous key papers on chemical evolution models of galaxies with
different types 
(e.g., Matteucci \& Francois 1989; Pagel \&  Tautvai\v{s}ien\'{e} 1998;
Lanfranchi \&  Matteucci 2010; Kirby et al. 2011; Tsujimoto \& Bekki 2012),
we do not discuss each of these in detail, because it is simply beyond
the scope of this paper. 
We do not discuss the origin of intriguing chemical abundances of dwarfs 
(Sgr and ultra-faint dwarfs, e.g.,  McWilliam et al. 2013; Roederer \& Kirby 2014) 
other than the LMC and Fornax either in this paper.

\section{The model}

\subsection{Outline}
This paper describes our first attempt to investigate 
the possible influences of radiation-driven
dust wind on galactic chemical evolution.  We therefore
adopt a rather idealized one-zone chemical evolution
model in order to demonstrate such influences more clearly.
The present study adopts the following big picture of dust removal processes in galaxies.
First, gas and dust mostly in cool ISM
can be expelled from disk to halo regions of a galaxy through
energetic stellar winds and radiation pressure of stars on dust grains.
At this stage,  dust and gas may or may not be hydrodynamically coupled with each other
depending on the gas densities of the ejected matter and the relative velocity
of gas and dust.
Then radiation pressure on dust can expel only  the dust further from the halo region
so that dust can be completely removed from the galaxy and thus can not be recycled
into the original ISM. This removal process of metals through
dust wind is different in different heavy elements
(Mg, Ca, and Fe) so that the time evolution of chemical abundance patterns can be
significantly influenced by the removal process.

The present one-zone model
adopts three assumptions on (i) from where dust can be removed more efficiently
in gas disks (e.g., from cool or warm ISM), (ii) whether dust and gas dynamics
can be coupled with each other,
and (iii) how  the removal processes 
of  heavy metals (e.g., Mg, Ca, and Fe) can depend on
the depletion levels and  sizes  of dust containing the metals, 
and they are described below in \S 2.1.1-3.
The present  big picture of the dust removal process from galaxies
is based on the three assumptions
in the present study.
We admit that the present models might be less realistic
in some points owing to these assumptions, but we consider that the present models
are reasonable enough to grasp some essential influences of dust wind
on chemical evolution of galaxies  in this first investigation.
More sophisticated models for dust removal processes should be constructed
in our future studies.

\subsubsection{Efficient dust removal from cool ISM} 

We assume that a much larger amount of dust can be removed from cool ISM rather than
warm and hot ISM in a galaxy.
This assumption is quite reasonable as follows.
Previous observations revealed that dust depletion for heavy elements (Fe, Mg, and Ca)
is much more severe in cool ISM than in warm and hot ISM (e.g., Welty et al. 1999).
For example, the Ca-depletion level in cool ISM is $\sim 16$ times greater than
that in warm ISM (See Fig. 6 in Welty et al. 1999). Furthermore,
the typical  hydrogen number fraction per unit volume 
($f_{\rm H} \propto f_{\rm V} n_{\rm H}$, where $f_{\rm V}$ and $n_{\rm H}$ are the volume
filling factor and hydrogen number density at each phase, respectively) 
in the hot, warm, and cool (cool and diffuse
H$_2$ and cool H~{\sc i}) 
are 0.002, 0.2, and 0.6 cm$^{-3}$, respectively (Drain 2009).
These observations  mean that the vast majority of heavy metals
(that are relevant to the present study) are locked up in dust of cool
ISM. In the three-component ISM (McKee \& Ostriker 1977),
the total mass of metals removed from a 
galaxy ($\delta M_{\rm metal}$) through dust wind can be given as follows:
\begin{equation}
\delta M_{\rm metal} 
\propto \delta (M_{\rm dust, hot}+M_{\rm dust, warm}+M_{\rm dust, cool}),
\end{equation}
where $M_{\rm dust, hot}$, $M_{\rm dust, warm}$, and $M_{\rm dust, cool}$ are the 
total dust masses in hot, warm, and cool ISM, respectively.
By considering the above observations, we assume that
\begin{equation}
\delta M_{\rm metal} \propto  \delta M_{\rm dust, cool}.
\end{equation}
We accordingly use the observed dust depletion patterns of individual elements (e.g., Mg
and Ca) for cool ISM in order to estimate the total amount of metals that are locked
up in dust and can be thus removed from a galaxy through radiation-driven dust wind.
The adopted dust depletion levels for the investigated
elements in this study is summarized in Table 1.
The detailed model for this dust removal process is given later in \S 2.3.

Since depletion levels are rather high in cool ISM and they are quite different
between different elements, selective removal of dust from cool ISM can influence galactic
chemical evolution and thus is  worthy of a detailed investigation.
However,  even if dust is removed selectively from warm and hot ISM,
such removal can not influence the time  evolution of abundance 
patterns in galaxies significantly
owing to the rather low depletion levels and the smaller element-to-element differences
in the depletion levels.
An idealized model for the adopted selective loss of dust from cool ISM is presented
later in this paper.

\subsubsection{Dust-gas decoupling in galactic halos}

We assume that only dust
can be removed completely from a galaxy (i.e., from its  halo)
through radiation pressure of stars for some physical conditions
so that dust can not be recycled later into ISM of the galaxy
(i.e., gas can not be escaped from the galaxy).
This assumption can be reasonable and realistic as follows.
Many authors have already investigated physical conditions
for dust-gas hydrodynamical coupling in galaxies
(e.g., Spitzer 1978;  Franco et al. 1991; F91;
Davies et al. 1998),
and found that dust-gas coupling is possible in the relatively high-density part of ISM
in galaxies.
For example,
Franco et al. (1991) shows that gas clouds
can be transferred to high latitudes in the Galaxy, because radiation pressure
on dust grains can raise both the dust and gas above the gas disk owing to the
dust-gas coupling ('photolevitation process').

However, F91 demonstrated that after dust can be located above
the disk owing to the photolevitation process,   the dust can be further
expelled by radiation wind from a galaxy (depending on several parameters though).
These previous theoretical models
suggested that dust can be selectively removed from galaxies
to become intergalactic dust, though the final states of dust depend strongly on
the details of their models.
More recently, Murray et al. (2005) and Coker et al. (2013) discussed dust-gas coupling
in galaxies and found that dust-gas coupling can be important on galaxy-scale gas
dynamics such as dust wind evolution
only if ISM density can be as high as or higher than $0.01-0.1$ cm$^{-3}$. This means
that in the low-density halo regions 
(less than $\sim 10^{-4}-10^{-5}$ cm$^{-3}$ 
for the Galaxy, e.g., Sembach et al. 2003)
of galaxies, gas and dust should be dynamically decoupled.
It is therefore 
highly likely that only dust can escape from galaxies (if radiation pressure
is strong enough) after both dust and gas have reached  the  halo regions.

The assumed selective removal of dust is consistent with recent observational results
by Xilouris et al. (2006), who found a very large dust-to-gas-ratio ($D \sim 0.05$,
which  is about 6 times larger than the solar neighborhood) in the outer halo of M81.
This can not be explained if both gas and dust are removed equally from
galaxies in M81 group and later located in the outer halo region of M81.
The observed very large dust-to-gas ratio can be explained, if dust can
be much more efficiently removed from galaxies than gas
to reach the outer halo region of M81 finally.
Furthermore, such selective removal of dust from galaxies appears to be
consistent with a recent observation that about 50\% of
all dust is located in outer halo regions (at $R \sim 20$ kpc
to several Mpc) of luminous  galaxies (M\'enard et al. 2010).
We therefore consider that much more efficient removal of dust from galaxies
(in comparison with gas) is quite reasonable
and realistic (i.e., consistent with observations) 
in the present study.

\subsubsection{Dust removal efficiencies dependent  on dust depletion levels}

As shown by F91, the dust removal efficiencies depend on  dust sizes 
for a given time-dependent radiation field of stars in a galaxy.
Therefore, if dust sizes are  different between different dust populations
(e.g., Mg-bearing and Ca-bearing dust),
then the removal efficiencies could be different between them. 
Since we do not have
enough observational details on the size/composition differences in different dust
populations, we can not currently construct a realistic model for
dust size distributions for individual dust populations based on observations.

Theoretical models on dust properties of stars have provided
dust compositions just for a number of key individual dust populations 
(e.g., MgSiO$_4$, MgO, Al$_2$O$_3$,
and FeS)
in AGB stars and SNe (e.g., Nozawa et al. 2003; Piovan et al. 2011).
Since the detailed information on the
size distributions of  Mg- and  Ca-bearing dust (that are
important in the present study) have not been provided yet,
we can not currently investigate the possible
differences in dust removal efficiencies between
different dust populations (owing to  different responses of dust to radiation
field between these dust caused by their  different size distributions and compositions)
in a quantitative manner.
However, we can discuss the possible differences in the sizes between
Mg- and Ca-bearing dust based on some results of recent theoretical works on
dust size distributions.

Nozawa et al. (2003) investigated the dust size distributions produced by
SNe with different masses for a number of
dust population (e.g., Mg$_2$SiO4 and FeS) and found that the typical dust sizes
range from  $2 \times 10^{-4} \mu$m  to $10^{-1} \mu$m (See their Fig. 10). 
The dust produced by AGB stars are demonstrated to have size distributions
biased toward the larger size  of $\sim 0.1\mu$m  (e.g., Winters et al. 1997; 
Yasuda \& Kozasa 2012).
These theoretical results suggest  that the typical size of SNe dust 
is significantly smaller than that of AGB dust.

Mg-bearing dust is demonstrated to be  formed efficiently both in AGB stars 
(e.g., Ferrarotti \& Gail 2006) and SNe (e.g., Nozawa et al. 2003)
while the formation efficiency of Ca-bearing dust (CaCO$_3$) is suggested
to be negligibly small in AGB stars (Ferrarotti \& Gail 2005).
Therefore, if Ca-bearing dust can be more preferentially formed in SNe,
the above theoretical results by Nozawa et al. (2003) and Winters et al. (1997)
would imply that
the typical size of Ca-bearing dust can be smaller than that of Mg-bearing dust.
If the sizes of Ca-bearing dust are really systematically smaller than
those of Mg-bearing dust, then the Ca-bearing dust can be more efficiently
removed from a galaxy  for a given radiation field of stars in  the galaxy,
because smaller dust grains can be more efficiently removed 
from galaxies by radiation pressure (F91).

It should be stressed, however, that the later evolution of dust by destruction
processes of SNe and dust coagulation processes
in ISM can significantly change the dust size distributions (e.g., Hirashita \& Yan 2009).
It is therefore reasonable for us to consider that
we simply do not know the typical sizes  of Mg- and Ca-bearing dust in real ISM.
It is ideal that we can adopt a model in which the removal efficiency of a ($i$-th) metal
element ($\epsilon_i$) through dust wind for a given radiation field 
depends both on the typical size of dust containing the element ($\lambda_i$)
and the dust depletion level of the element ($\delta_i$) as follows:
\begin{equation}
\epsilon_i=F(\lambda_i,\delta_i),
\end{equation}
where the functional form ($F$) should be modeled properly.
However, owing to the lack of observational and theoretical works on
$\lambda_i$ for heavy elements relevant to the present work (e.g., Mg, Ca, and Fe),
we can not include the dependences of $\epsilon_i$ on $\lambda_i$
in the present study 
and thus we assume that $\epsilon_i$ depends solely on $\delta_i$ as follows:
\begin{equation}
\epsilon_i=F(\delta_i).
\end{equation}
We admit that this model is oversimplified to some extent and thus suggest that
the present results could be changed if a realistic model for
the influences of the typical sizes of different dust populations
on dust removal efficiencies is included.
We briefly discuss the possible influences of dust size differences on the present
results later in \S 4.1.

Thus we here assume that the removal efficiency
of an element through dust wind
depends solely on the dust depletion level for a given radiation field of a galaxy.
The details of the model are given later in \S 2.3.
For the adopted model,  a larger mass fraction of Ca is locked up
in dust (than Mg) so that a larger amount of Ca metal
can be efficiently removed from a  galaxy
through dust wind than Mg.
This is quite reasonable, because Ca in cool ISM of the Galaxy is observed
to be more severely dust-depleted than
Mg (i.e., a larger amount of metal is locked up in dust for Ca).  However,  it should
be noted that we do not know direct evidence for this more efficient loss of Ca
in real galaxies.  What we can do in this paper is to investigate in what models
for dust removal efficiencies
the observed abundance patterns of galaxies can be better reproduced.
We thus fully admit that there would be an uncertainty in the present model for
the possibly different dust removal efficiencies between different elements.

\begin{table}
\centering
\begin{minipage}{80mm}
\caption{The depletion level ($\delta_i$) for the selected four elements.}
\begin{tabular}{cc}
{  Element  \footnote{
The time evolution of these four elements
are investigated in detail by the present one-zone chemical
evolution models.
}} &
{ $\delta_i$   \footnote{
These values are calculated for the
data given in SS96 and Draine 2009. The smaller $\delta_i$ for the $i$-th element
means that a larger amount of the element
is locked up in dust grains.
}}  \\
Mg & $1.1 \times 10^{-1}$   \\
Ca & $1.7 \times 10^{-4}$   \\
Fe & $3.7 \times 10^{-3}$   \\
Ti & $1.0 \times 10^{-3}$   \\
\end{tabular}
\end{minipage}
\end{table}

\begin{table}
\centering
\begin{minipage}{80mm}
\caption{Model parameters for one-zone chemical evolution.}
\begin{tabular}{cccc}
{  Model  \footnote{
The `M', `F', and `E'  are referred to as
the Magellanic Clouds (LMC), Fornax, and Early-type galaxy models, respectively.
The models with no values of $\beta$, $\gamma$, and $C_{\rm r}$ indicated
(e.g., M1 and F1) are those
without dust wind.
}} &
{  $\beta$  \footnote{
The parameter that controls dust removal fraction and defines the
minimum level of dust removal.
}} &
{ $\gamma$   \footnote{
The parameter that controls dust removal fraction and determine
the degree of differential dust removal among different elements.
}} &
{ $C_{\rm r}$   \footnote{
The parameter that controls the strength of radiation pressure
of stars on dust grains.
}} \\
M1 & ... & ... & ...  \\
M2 & 0 & $-1$  & $10^{-3}$    \\
M3 & 0.1 & $-1$  & $10^{-3}$    \\
M4 & 0.3 & $-1$  & $10^{-3}$    \\
M5 & 0.5 & $-1$  & $10^{-3}$    \\
M6 & 0.3 & $-0.5$  & $10^{-3}$    \\
M7 & 0.3 & $-0.3$  & $10^{-3}$   \\
M8 & 0.1 & $-0.5$  & $10^{-3}$    \\
M9 & 0.1 & $-0.3$  & $10^{-3}$    \\
M10 & 0 & $-0.5$  & $10^{-3}$    \\
F1 & ... & ... & ...    \\
F2 & 0 & $-1$  & $6 \times 10^{-3}$   \\
E1 & ... & ...  & ...    \\
E2 & 0 & $-1$  & $10^{-3}$   \\
E3 & 0 & $-1$  & $7 \times 10^{-4}$    \\
E4 & 0 & $-1$  & $3 \times 10^{-4}$    \\
\end{tabular}
\end{minipage}
\end{table}

\begin{figure*}
\psfig{file=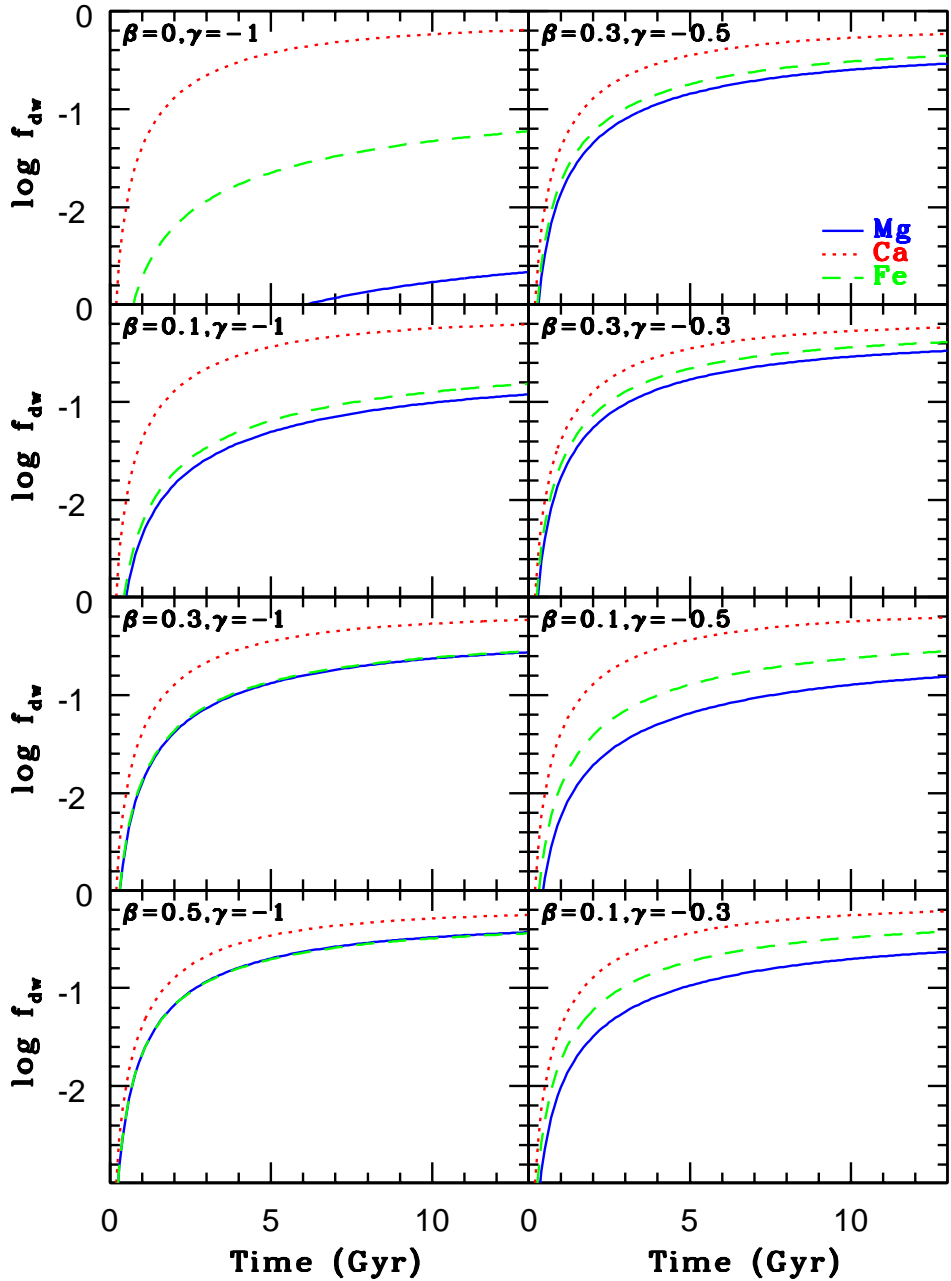,width=14.0cm}
\caption{
The time evolution of dust removal fraction ($f_{\rm dw, \it i}$) for $i=$ Mg
(blue solid), Ca (red dotted), and Fe (green short-dashed) in the LMC dust wind
models with different $\beta$ and $\gamma$ indicated in the upper left panel
of each frame. The larger $f_{\rm dw, \it i}$ means that a larger fraction
of dust (thus metal) is removed from the main body of a galaxy through
radiation-driven dust wind.
}
\label{Figure. 1}
\end{figure*}

\subsection{Basic equations}

We adopt one-zone chemical evolution models that are essentially the same
as those adopted in our previous studies on the chemical
evolution of the LMC (BT12).
Accordingly, we briefly describe the adopted models
in the present study.
A galaxy  is assumed to form through a continuous
gas infall from outside the disk region (e.g., halo) for $0 \le t \le t_{\rm end}$,
where $t=0$ is the starting time of a calculation. 
The final time  $t_{\rm end}$ of a calculation is set to be 13 Gyr for most models.
We investigate the time evolution of the gas mass fraction
($f_{\rm g}(t)$), the star formation rate ($\psi(t)$), and
the abundance of the $i$th heavy element ($Z_i(t)$) for a given
accretion rate ($A(t)$), IMF,
ejection rate of ISM due to SNe ($w(t)$) 
and ejection rate of ISM due to SNe ($w_{\rm d} (t)$) {\it through dust wind.}

The basic equations for the adopted one-zone chemical evolution models
are described as follows:
\begin{equation}
\frac{df_g}{dt}=-\alpha_{\rm l}\psi(t)+A(t)-w(t)-w_{\rm d}(t)
\end{equation}
\begin{eqnarray}
\frac{d(Z_if_g)}{dt}=-\alpha_{\rm l} Z_i(t)\psi(t)+Z_{A,i}(t)A(t)+y_{{\rm II},i}\psi(t) 
\nonumber \\  +y_{{\rm Ia},i}\int^t_0 
\psi(t-t_{\rm Ia})g(t_{\rm Ia})dt_{\rm Ia}
\nonumber \\ 
-W_i(t) -W_{\rm d, \it i}(t) \ \ ,
\end{eqnarray}
\noindent where $\alpha_{\rm l}$ is the mass fraction 
(per unit mass) locked up in dead stellar remnants and long-lived stars, 
$y_{\rm {Ia}, i}$, $y_{\rm {II}, i}$
are the 
chemical yields (per unit mass) for the $i$th element from type II supernovae (SN II), 
from SN Ia, 
respectively, 
$Z_{A,i}$ is the abundance of heavy elements  contained in the infalling gas,
$W_i$ is the rate of supernova wind for each element,
and $W_{\rm d, \it i}$ is the dust wind rate for each element
(i.e., $w_{\rm d}$ is the sum of $W_{\rm d, \it i}$ for all elements).
Unlike BT12, we did not include the AGB term in the equations (5) and (6),
because (i) the present study does not investigate the chemical 
evolution of [Ba/Fe] and (ii) 
AGB feedbacks do not play a role in the time 
evolution of [Mg/Fe] and [Ca/Fe].

In the present study, the total mass of a galaxy ($M_{\rm t}$) including
both gas and stars is adopted
as mass units for all models and accordingly the masses of all components
(e.g., gas, stars, and metals)
are normalized by $M_{\rm t}$.
The gas mass fraction ($f_{\rm g}$) of a galaxy is normally defined as
the ratio of gas mass ($M_{\rm g}$) to $M_{\rm t}$.
The dust wind rate ($w_{\rm d}$) is defined as the rate of mass ejection
by radiation-driven wind per unit mass (i.e., normalized by $M_{\rm t}$).
The star formation ($\psi(t)$)  and accretion rates ($A(t)$)
are also those per unit mass
(i.e., they are normalized SFRs and accretion rates).
Unlike our previous studies (BT12), we do not include the ejection of metals 
through stellar wind (i.e., $w(t)=0$),  mainly because we try to clearly understand the roles of
dust wind in galactic chemical evolution. Furthermore, as discussed in BT12,
the models with no stellar wind can reproduce the observed age-metallicity relation
and chemical abundances of the LMC by choosing a reasonable IMF.
All of the elements in the stellar ejecta from SNe 
are assumed to be returned back to
ISM of galaxies immediately after SN explosions.

The above chemical yields are defined as the masses of newly synthesized
metals per unit mass  and thus are dimensionless quantities.
The chemical abundance of an element at each time is normally defined as the 
ratio of the total metal mass ($M_{\rm Z, \it i}$) of the element
to the total gas mass ($M_{\rm g}$)
at the time. 
Both $W_i$ and $W_{\rm d, \it i}$ are wind rates per units mass (i.e., normalized
by $M_{\rm t}$).
It should be stressed  here again that $W_i$ is set to be 0 for all elements
so that we can discuss more clearly the roles of dust wind in galactic chemical evolution.

The quantity $t_{\rm Ia}$  represents
the time delay between star formation and SN Ia explosion.
The term $g(t_{\rm Ia}$) is the distribution
functions of SNe Ia,
and the details of which  are described later
in this section. 
Thus equation (5) describes the time evolution
of the gas due to star formation, gas accretion, and dust wind.
Equation 6 describes 
the time evolution
of the chemical abundances due to chemical enrichment by supernovae
and metal removal through dust wind.
Unlike BT12, the present study  does  not investigate the chemical evolution of [Ba/Fe],
and thus the AGB term in the equations used in BT12 is not included in the above equations
(5) and (6).

\subsection{Dust wind rate}

The dust wind rate $W_{\rm d, \it i}(t)$ 
is a function of (i) a probability ($P_i$) that $i$-th element (e.g., Mg)  can be locked up in
dust grains and influenced by stellar radiation and (ii) the strength of stellar radiation ($S(t)$) that
can be exerted on dust grains. Therefore $W_{\rm d, \it i}$ is described as follows:
\begin{equation}
W_{\rm d, \it i}(t)=P_i S(t).
\end{equation}
The probability $P_i$ is assumed to depend solely on the depletion level ($\delta_i$)
for the $i$-th element
and described as follows:
\begin{equation}
P_i=\beta+C_{\rm p}(1-\beta)\delta_i^{\gamma},
\end{equation}
where $\beta$ and $\gamma$ are the two basic parameters that control the dependences
of $P_i$ on $\delta_i$ and $C_{\rm p}$ is a normalization constant. 
In this power-law formula, the larger absolute magnitude of $\gamma$ means
a steeper (stronger) dependence of $P_i$ on dust depletion levels.
$\beta$ defines the minimum possible $P_{\rm i}$ for each element,
because $\beta$ is assumed to be equal to or larger than 0.

We adopt this simplified power-law 
formula for $P_i$,  because the present work is 
the very  first study on the effects
of dust wind on chemical evolution of galaxies and thus
needs to demonstrate the basic dust effects more clearly.
The power-law formula would make it more straightforward for us 
to interpret the present results of the models with different model parameters
(e.g., $\beta$ and $\gamma$).
It should be noted here that there is currently no/little
strong observation constraint
on how $P_i$ might depend on dust properties (e.g., depletion levels
and dust sized etc). Therefore the adopted formula for $P_i$ might be regarded
as an arbitrary one. However, we think that we can more clearly
understand  the essential ingredients of the dust wind effects on galactic
chemical evolution
thanks  to the relatively simplified formula for $P_i$ in the present
study.

The depletion level $\delta_i$ is defined as follows:
\begin{equation}
\delta_i=\frac{ (N_{i, \rm g}/N_{\rm H}) }{ (N_i/N_{\rm H})_{\odot} },
\end{equation}
where $(N_{i, \rm g}/N_{\rm H})$ is the gas-phase abundance for the $i$-the element
(relative to H) and $(N_i/N_{\rm H})_{\odot}$ is the corresponding solar value.
Accordingly, the lower $\delta_i$ means the higher level of dust depletion.
Since almost all (99.98\%) of Ca is dust-depleted,  the constant $C_{\rm p}$ is chosen
such that $P_i$ for Ca can be 1 for a given $\beta$ and $\gamma$.
In the above formula for $P_i$, $\beta$ should range from 0 to 1 while $\gamma$ should be
negative. We consider that the model with $\beta=0$ and $\gamma=-1$ can be a reasonable combination
in the present study, because the dust removal efficiency is proportional to
the dust depletion level.
We however investigate models with different $\beta$ and $\gamma$, because
these two parameter are not observationally constrained.
We adopt the observed values of $(N_{i, \rm g}/N_{\rm H})$  and $(N_i/N_{\rm H})_{\odot}$
from Table 23.1 in Draine (2009).

\subsection{Radiation pressure of stars}

The strength of radiation pressure of stars is assumed to be proportional to the total luminosity
of a galaxy as follows:
\begin{equation}
S(t) = C_{\rm r}\int^t_0 
f_{\rm M/L}^{-1}(t-T) M_{\rm ns}(T) dT,
\end{equation}
where $f_{\rm M/L}$ is the mass-to-light ratio ($M/L$) of a single stellar population (SSP)
and $M_{\rm ns}(T)$ is the total mass of stars formed at $t=T$.
Since the $M/L$ of a SSP is a function of age and metallicity, we need to use a stellar population
synthesis code in order to properly calculate $f_{\rm M/L}$ at each time step.
Accordingly,  we use the code MILES  (Vazdekis et al. 2010) for the M/L estimation of all models
in the present
study.
The parameter $C_{\rm r}$ can control how much amount of metals can be removed through
radiation-driven wind, and it is fixed at $10^{-3}$ for most models. For this value of $C_{\rm r}$
in the LMC model later described, a significant fraction of metals can be removed so that
chemical evolution can be influenced by dust wind.

\begin{figure*}
\psfig{file=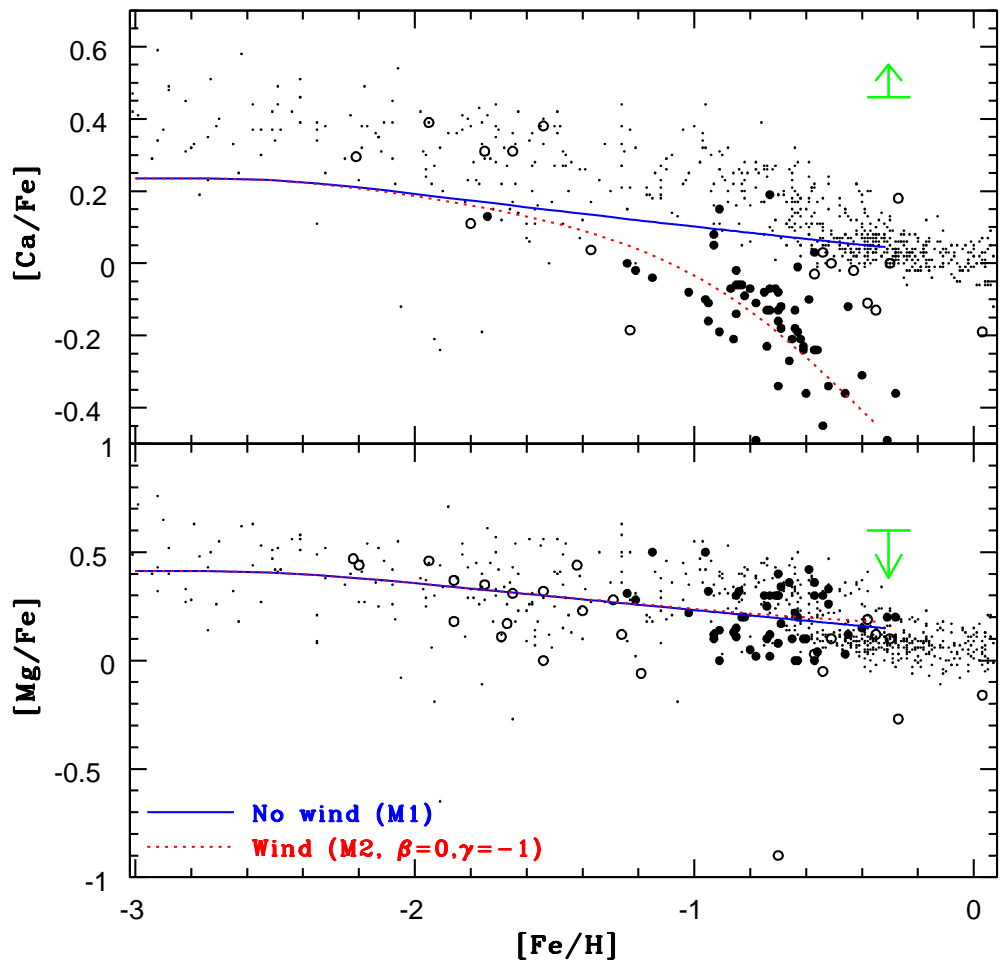,width=14cm}
\caption{
The evolution of [Ca/Fe] (upper) and [Mg/Fe] (lower) as a function of [Fe/H]
for the LMC model without dust wind (M1, blue solid) and with dust wind (M2, red dotted).
The observed locations of the  LMC field stars (big filled circles)
and clusters (big open circles) and the Galactic field stars
(small dots) on the [Mg/Fe]-[Fe/H] plane  are shown for comparison.
The observational results  include
P08
for the LMC field stars,
Johnson et al. 2006, Mucciarelli et al. (2008, 2010, 2011),  and C12
for the LMC clusters,
and Venn et al. (2004)
for the Galactic field stars.
The green arrows indicate the possible differences in the observed [Ca/Fe]
(+0.09 dex) and [Mg/Fe] ($-0.22$ dex) between P08 and V13 (i.e.,
the most recent observational study).
The higher [Ca/Fe] in V13 implies that if the models
are compared with V13, then weaker dust wind is required
to reproduce the observed [Ca/Fe] at higher metallicities in the LMC.
}
\label{Figure. 2}
\end{figure*}

\begin{figure*}
\psfig{file=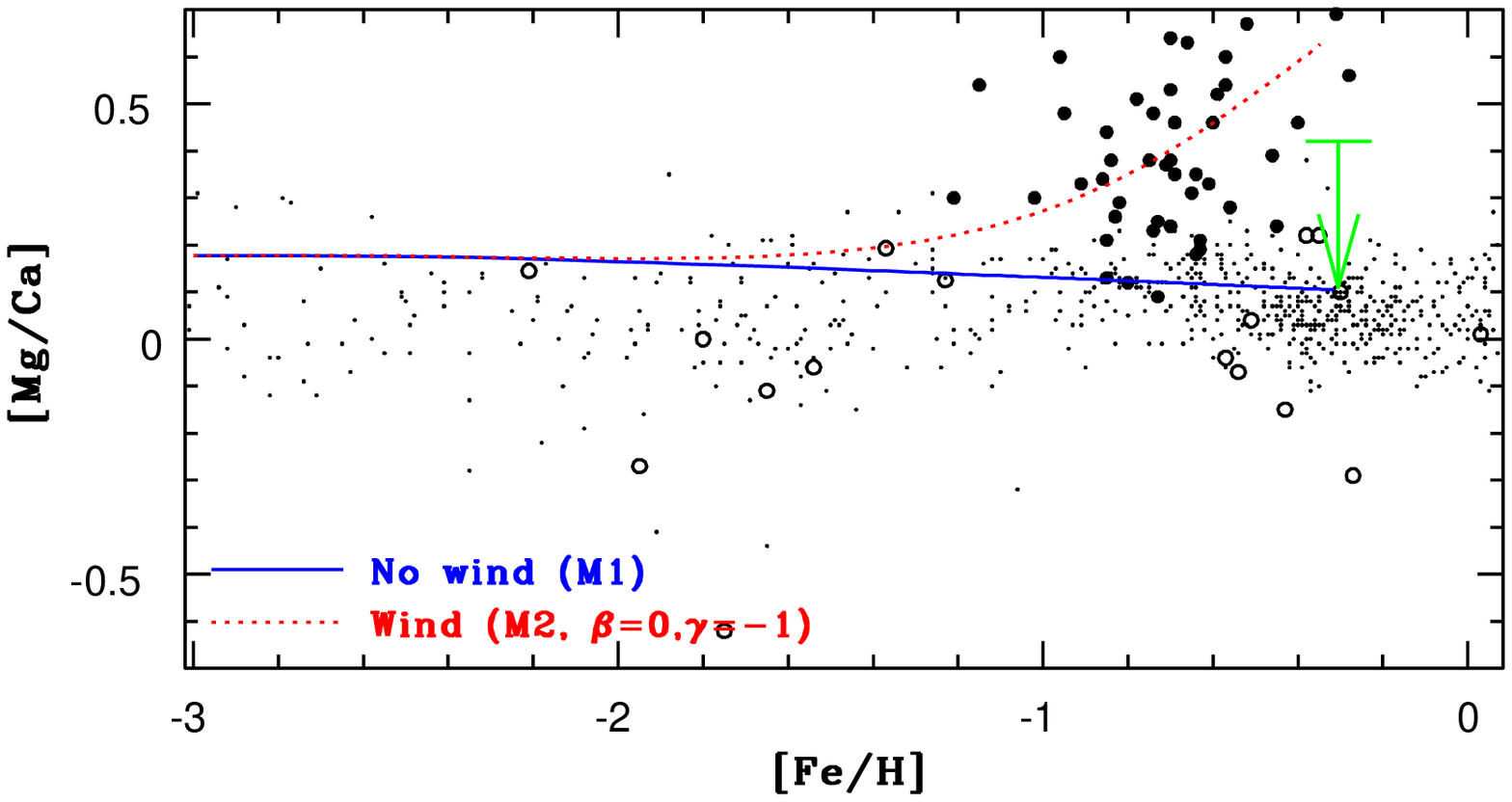,width=14cm}
\caption{
The evolution of [Mg/Ca] as a function of [Fe/H]
for the LMC model without dust wind (M1, blue solid) and with dust wind (M2, 
red dotted).
The green arrow indicates the possible difference in the observed [Mg/Ca]
($-0.31$ dex) between P08 and V13. 
}
\label{Figure. 3}
\end{figure*}

\subsection{Star formation and gas accretion rate}

The star formation rate per unit mass ($\psi(t)$)
is assumed to be proportional
to the gas fraction with a constant star formation
coefficient and thus is described as follows: 
\begin{equation}
\psi(t)=C_{\rm sf}f_{\rm g}(t)
\end{equation}
The star formation coefficient ($C_{\rm sf}$) can control the rapidity
of gas consumption and $C_{\rm sf}=0.006$ is the reasonable value for the LMC
(BT12) for a given set of reasonable model parameters.
For the models with $C_{\rm sf}=0.006$,  both the typical SF rate and
the final total stellar mass ($M_{\rm s}=2.7 \times 10^{9} M_{\odot}$)
observed in the LMC
can be reproduced by some models (BT12).
For the accretion rate, we adopt the formula in
which $A(t)=C_{\rm a}\exp(-t/t_{\rm a})$
and $t_{\rm a}$ is a free parameter controlling the time scale of 
the gas accretion.   The normalization factor $C_{\rm a}$ is determined such
that the total gas mass accreted onto a galaxy  can be 1 (in model units) for a given
$t_{\rm a}$. We investigated  models with different $t_{\rm a}$ for different
three types of models later described.
The initial [Fe/H] of the infalling  gas is set to be $-3$  and
we assume a SN-II like enhanced [$\alpha$/Fe] ratio
(e.g., [Mg/Fe]$\approx 0.4$).

\subsection{Chemical yields and delay time distribution of SN Ia}

We adopt  the nucleosynthesis yields of SNe II and Ia from Tsujimoto et al. 1995 (T95)
to deduce $y_{{\rm II},i}$ and $y_{{\rm Ia},i}$ for a given IMF.
Stars with masses larger than $8 M_{\odot}$ explode as SNe II soon after
their formation and eject their metals into the  ISM.
In contrast,  there is a time delay ($t_{\rm Ia}$) between the star formation
and the metal ejection for SNe Ia. We here adopt the following delay time distribution
($g(t_{\rm Ia}$)) for 0.1 Gyr $\le t_{\rm Ia} \le$ 10 Gyr,
which is consistent with recent observational studies
on the SN Ia rate in extra-galaxies (e.g., Totani et al. 2008; Maoz et al. 2010,
2011):
\begin{equation}
g_{\rm Ia} (t_{\rm Ia})  = C_{\rm g}t_{\rm Ia}^{-1},
\end{equation}
where $C_{\rm g}$ is a normalization constant that is determined by
the number of SN Ia per unit mass  (which is controlled by the IMF
and the binary fraction for intermediate-mass stars
for  the adopted power-law slope of $-1$).
The fraction of the stars that eventually
produce SNe Ia for 3--8$M_{\odot}$ has not been observationally determined
and thus is regarded as a free parameter, $f_{\rm b}$.
We mainly
investigate models with different$f_{\rm b}=0.03$ and 0.09 for three different types
of models later described. 

Like BT12, we use the theoretically predicted yields of SN explosions from T95
for consistency between BT12 and the present work.  
Accordingly,  the initial [Ca/Fe] is $\sim 0.24$
that are slightly
lower than the observed [Ca/Fe] ($0.3 \sim 0.4$) for the Milky Way (MW) halo stars. 
We here stress that the key conclusion on the origin
of the observed low [Ca/Fe] of the LMC stars in this paper remains unchanged, if we 
adopt different  Ca yields.  We discuss this point briefly in the Appendix A by using
the results of the models with different Ca yields.

\subsection{IMF}

The adopted IMF  is  defined
as $\Psi (m_{\rm I}) = M_{s,0}{m_{\rm I}}^{-\alpha}$,
where $m_{\rm I}$ is the initial mass of
each individual star and the slope $\alpha =2.35$
corresponds to the Salpeter IMF  (Salpeter 1955).
The normalization factor $M_{s,0}$ is a function of $\alpha$,
$m_{\rm l}$ (lower mass cut-off), and  $m_{\rm u}$ (upper mass cut-off).
These  $m_{\rm l}$ and $m_{\rm u}$
are  set to be   $0.1 {\rm M}_{\odot}$
and  $50 {\rm M}_{\odot}$, respectively (so that the normalization
factor $M_{s,0}$ is dependent simply on $\alpha$).
We investigate models with different $\alpha$ to find the  model(s)
that can best explain the observed abundance patterns of stars in the LMC.
We do not discuss models with
different $m_{\rm u}$,  because the effects of changing
$m_{\rm u}$ on the LMC chemical evolution
are similar to those of changing $\alpha$.

As shown in BT12, a steeper IMF ($\alpha=2.55$) can slightly better
explain the observed chemical abundances and gas mass fraction of the LMC
than the Salpeter IMF, if  stellar wind (from SNe)
is not included in the models. Since the main purpose
of this paper is to demonstrate the dust wind effects on galactic 
chemical evolution clearly (not to find the best IMF model),
it would be important for this paper to demonstrate that the  dust wind 
effects derived from the Salpeter IMF model
can be true for other models with different IMFs, in particular,
$\alpha=2.55$.
We thus briefly discuss the dependences of the present results on IMFs
in the Appendix B.

Recent numerical simulations have clearly shown that galaxy-scale stellar
winds driven by SN explosions can eject a significant amount of ISM out of dwarf
galaxies (e.g., Recchi et al. 2013; Ruiz et al. 2013), which implies 
that the present model assumption of no stellar wind would be less
realistic for the LMC. Although BT12 have already investigated
the effects of stellar winds on the chemical evolution of the LMC,
it would be important for this study to confirm that
the dust wind effects on galactic chemical evolution derived from
the models with no stellar winds can be true 
also for the models with stellar winds.
Since the detailed investigation of stellar wind effects on galactic
chemical evolution is not the major purpose of this paper,
we discuss this issue by using a reasonable set of the LMC models
with stellar winds in the Appendix C.

\subsection{Main points of analysis}
\subsubsection{Comparison between observations and models}

We discuss how dust wind influences chemical evolution of galaxies by comparing
the results of the present chemical evolution models with recent observations derived
for the LMC. We therefore adopt a reasonable set of model parameters for the LMC (BT12)
and thereby investigate the time evolution of several selected elements (e.g. Fe, Ca, and Mg).
We mainly investigate the LMC because of the two reasons described in \S 1.
Like BT12, we focus on the observational results by P08, which clearly shows intriguing
 results on 
the [Ca/Fe]-[Fe/H] and [Mg/Ca]$-$[Fe/H] relations of the LMC.
The unusually low [Ca/Fe] ($<-0.3$) and high [Mg/Ca] ($>0.3$) 
for stars that are not so metal-poor in P08 were found to be hardly
explained by our previous models with a standard set of model parameters for the IMF and the
star formation history of the LMC.

However, it should be noted here that 
the latest results by V13 show a higher [Ca/Fe] (by 0.09 dex) 
in comparison with P08. This difference between the two observational studies
can be seen in other $\alpha$-elements (e.g., [Mg/Fe] by $0.22$ dex),
which implies that observational results need to be carefully compared
with theoretical ones.
These differences suggest 
that the inconsistency between our previous models
(BT12) and observations is not so large  in terms of 
[Ca/Fe]$-$[Fe/H] and [Mg/Ca]$-$[Fe/H] 
relations.  In the present study, we still mainly use the results by P08,
because we need to compare the previous models in BT12, which failed to
explain the results by P08, with the present new one, by using the same
observational data sets. This comparison allows us to demonstrate
the improvement of the present model over the previous one more clearly.
We later discuss whether  weaker dust wind is  required for explaining
the observational results by V13.

\subsubsection{Parameter study}

We mainly investigate the LMC model (M1-M10) with the Salpeter IMF, $f_{\rm b}=0.03$
$C_{\rm sf}=0.006$,  $t_{\rm a}=0.3$ Gyr in the present study. 
We choose these parameter values, because they can better reproduce the
age-metallicity relation and chemical abundance patterns in the LMC (BT12).
The parameters values
for dust wind in these models ($\beta$, $\gamma$, and $C_{\rm r}$) are different
and summarized in Table 2. We do not discuss how the IMF slope and $f_{\rm b}$ influence
the chemical evolution of galaxies,
because BT12 already discussed these in detail.  We also investigate the models 
that are reasonable for the Fornax dwarf galaxy (F1-F2)
and giant elliptical galaxies (E1-E4) in order to
discuss the origin of the observed intriguing abundances of these galaxies.
Since the results of the LMC are more important than other models (F1-F2 and E1-E4),
we first describe the results of the LMC model in \S 3. We describe the model parameters
for the Fornax galaxy and early-type galaxy models and discuss briefly the results of
the  models in \S 4. The parameter values for these models are summarized in Table 2.

Once metals are removed from the main body of a galaxy to be located
in the halo region through radiation-driven
dust wind,  the metals can not be returned back to the main body. 
The two key parameters for this radiation-driven dust 
(metal) removal process in a galaxy,
$\beta$ and $\gamma$, can determine the mass fraction of $i$-the element
(metal) that is removed
from the main body of a galaxy ($M_{\rm halo, \it i}$) to the total
mass of the element in the main body and the halo
($M_{\rm tot, \it i}$).
The mass fraction ($f_{\rm dw, \it i}$) is defined as follows:
\begin{equation}
f_{\rm dw, \it i}= \frac{ M_{\rm halo, \it i} }{ M_{\rm  tot, \it i} }.
\end{equation}
This `dust (metal) removal fraction' ($f_{\rm dw, \it i}$)
evolves with time ($t$) and
can be significantly different between different elements
owing to the adopted dependence of $P_i$ on $\delta_i$. 
In the present model,  dust removal means metal removal from the main bodies
of galaxies so that higher $f_{\rm dw, \it i}$ means that 
a smaller amount of $i$-the metal can remain in the main bodies.
We investigate
how $f_{\rm dw, \it i}$ depends on $\beta$ and $\gamma$ for a given set of
model parameters (e.g., $C_{\rm r}$ and $C_{\rm sf}$).

\begin{figure*}
\psfig{file=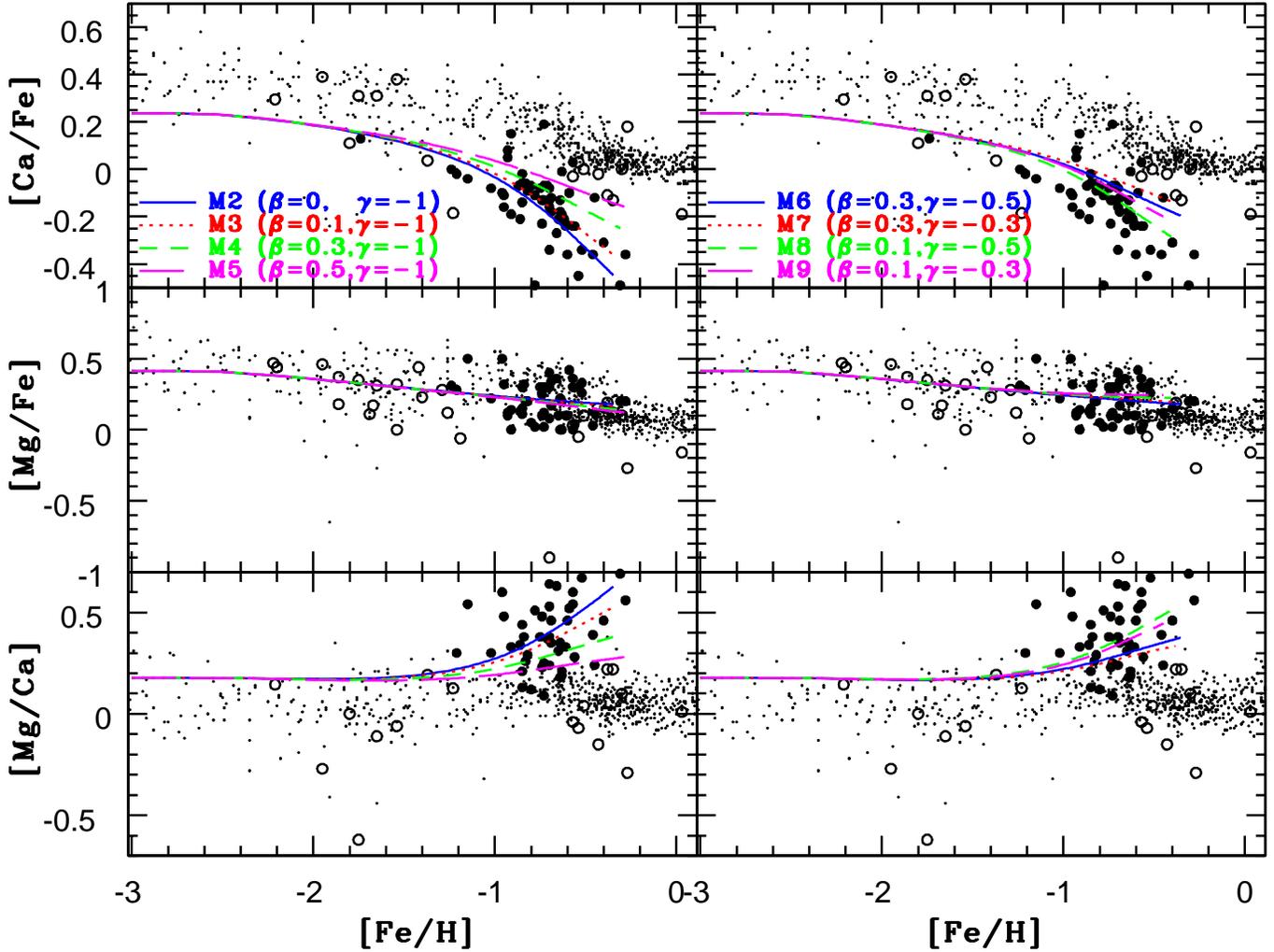,width=18cm}
\caption{
The evolution of [Ca/Fe] (top), [Mg/Fe] (middle), and [Mg/Ca] (bottom)
as a function of [Fe/H]
for the eight LMC models with different $\beta$ and $\gamma$.
In the left three frames,  blue, red, green, and magenta lines
represent the models with
$\beta=0$ and $\gamma=-1$ (M2),
$\beta=0.1$ and $\gamma=-1$ (M3),
$\beta=0.3$ and $\gamma=-1$ (M4),
and $\beta=0.5$ and $\gamma=-1$ (M5), respectively.
In the right three frames,  blue, red, green, and magenta lines
represent the models with
$\beta=0.3$ and $\gamma=-0.5$ (M6),
$\beta=0.3$ and $\gamma=-0.3$ (M7),
$\beta=0.5$ and $\gamma=-0.5$ (M8),
and $\beta=0.5$ and $\gamma=-0.3$ (M9), respectively.
}
\label{Figure. 4}
\end{figure*}

\section{Results}

\subsection{Evolution of $f_{\rm dw, \it i}$}

Figure 1 shows how $f_{\rm dw, \it i}$ evolves with time for three different
elements (Mg, Ca, and Fe) for a given set of dust wind parameters in the LMC
models with $C_{\rm r}=10^{-3}$. In this figure, the higher $f_{\rm dw, \it i}$
means a larger mass fraction  of the $i$-the element is removed 
from the main body of a galaxy. Clearly, Ca is the most efficiently removed
from the main bodies of galaxies for all 8 models (M1-M8) with different
$\beta$ and $\gamma$. The dust removal fraction of Fe is higher than that of
Mg at all time steps in the model M2 with $\beta=0$ and $\gamma=-1$
that could be the most reasonable in the present study.
The dust removal fraction between Fe and Mg can be clearly seen only
in the models with $\beta=0$ and 0.1 for a fixed $\gamma =-1$. 
The parameter $\beta$  for a fixed $\gamma$ can determine the minimum level
of dust removal fraction so that larger $\beta$ can cause the smaller differences
in dust removal fraction  between the three elements.

The differences in the dust removal fractions of Mg, Ca, and Fe are larger in
the models with $\gamma=-0.5$ than those with $\gamma=-0.3$ for a fixed $\beta$ (0.1
and 0.3). These differences in $f_{\rm dw, \it i}$ can cause the different evolution
in different elements, as described later. These models all show that the evolution
of $f_{\rm dw, \it i}$ is rather rapid in the early phase of galaxy evolution
($t<2$ Gyr) owing to the dramatic luminosity evolution caused by high SFRs.
However, the small $f_{\rm dw, \it i}$ (i.e., $\log f_{\rm dw, \it i}<-1$)
in this early-phase means that such dust removal can not so strongly influence the
chemical evolution of galaxies. 
In these LMC models,  metals in the main body of galaxies can be removed more slowly
yet steadily in the later evolution phase so that $f_{\rm dw, \it i}$ can be 
high ($>0.1$) enough to influence chemical evolution.
These results suggest that galactic chemical evolution can be more strongly influenced
by dust wind at their later evolution stages when galactic luminosities are high
enough to exert strong radiation pressure on dust grains.

\subsection{Correlations of [Mg/Fe], [Ca/Fe], [Mg/Ca] with [Fe/H]}

Fig. 2 compares the evolution of [Ca/Fe] and [Mg/Fe] as a function of [Fe/H]
in the two LMC models with (M2) and without dust wind (M1). Clearly, there is
no significant difference in the observed [Mg/Fe]$-$[Fe/H] relations of 
the LMC and the MW, though the LMC shows a large [Mg/Fe] scatter for a given [Fe/H]
at [Fe/H]$>-1$, which could be caused by secondary starburst due to tidal interaction
with the SMC (BT12). However, the [Ca/Fe]$-$[Fe/H] relation of the LMC is quite
different from that of the MW, in particular, for the higher [Fe/H] ($>-1$). 
These observations can not be well reproduced self-consistently by
the model M1 without dust wind. The models without dust wind can only explain
the observed [Mg/Fe]$-$[Fe/H] relation of the LMC by adopting a reasonable set of
model parameters, but it fails to explain the observed rather low [Ca/Fe] $<-0.3$ at
[Fe/H]$>-0.6$.

The dust wind (M2) can reproduce both the observed [Mg/Fe]$-$[Fe/H]
and [Ca/Fe]$-$[Fe/H] relations in a self-consistent manner for $C_{\rm r}=10^{-3}$,
$\beta=0$, and $\gamma=-1$.
As shown in Fig. 1,
Ca can be more efficiently removed from the main body of the LMC than 
Fe through radiation-driven dust wind in this  model M2. As a result of this,
[Ca/Fe] can decrease more rapidly  with increasing [Fe/H] in comparison
with the model M1 without dust wind.
Since both Mg and Fe are less efficiently removed from the main body of the LMC,
their time evolution can not be so strongly influenced by dust wind.
Therefore the [Mg/Fe]$-$[Fe/H] relation is not so different between the models
with and without dust wind. 
It should be noted here  that the apparently larger [Ca/Fe] differences at higher
[Fe/H] between the LMC and the MW can be reproduced reasonably well by the dust
wind model.

Fig. 3 shows that (i) the observed [Mg/Ca] are rather high ($>0.3$) for 
a sizable fraction of stars
in the LMC at [Fe/H]$>-1$ and (ii) [Mg/Ca] appears to increase with [Fe/H]
for the field stars of the LMC at [Fe/H]$>-1.2$.
Clearly, only the dust wind model can reproduce the observed rather high [Mg/Fe]
of some stars in the LMC at higher [Fe/H].
The apparent [Mg/Ca] increase with [Fe/H] can be well reproduced by the dust wind
model (M2) while the model M1 without dust wind
does not show such a trend at all.
These results imply that the observed trend is caused by a larger degree of
Ca removal through dust wind in the chemical evolution of the LMC.
However, the observed [Mg/Ca] dispersion at a given [Fe/H] is quite large, which makes
it difficult for the present study to propose a robust physical interpretation
of the observed apparent [Mg/Ca] increase with [Fe/H]. 
It is unclear why the observed [Mg/Fe] ([Ca/Fe]) of clusters in the LMC 
appear to be smaller (larger) than those
of the field stars. 

It should be stressed that the observed [Ca/Fe] by V13 is larger by 0.09 dex
than P08, which means that the observed  [Ca/Fe] at [Fe/H]$\sim -0.3$ is about
$-0.3$ rather than $-0.4$ in V13. Accordingly, if we adopt the V13 results, then
a weaker dust wind that removes a smaller amount of gas-phase metals
is required for the consistency of the model with the observation. 
The possible difference in [Mg/Ca] between P08 and V13 is 0.3 dex, which also
suggests that a significantly weaker dust wind is required to explain
[Mg/Ca]$\sim 0.3$ (rather than $\sim 0.6$) in the present models. 
Thus the required level of the removal process of Mg and Ca by dust wind
to explain the observed [Mg/Ca]$-$[Fe/H] relation
can be reduced, 
if the observed [Ca/Fe] and [Mg/Fe]
can be further systematically
higher and lower, respectively, in future observational studies.

It should be also noted that [Fe/H] is not so different between the two models
with and without dust wind, because gas-phase Fe is not removed so efficiently by
dust wind.
BT12 suggested that if Fe can be preferentially lost in the LMC
through some  wind processes, then the observed rather high [Ba/Fe]
would be explained , because  Ba is less efficiently
removed through stellar wind.
This implies that the present models with no stellar wind
would not  reproduce well the observed high [Ba/Fe] in the LMC,
even if the time evolution of [Ba/Fe] is investigated.

\subsection{Parameter dependences}

Fig. 4 compares the 
[Ca/Fe]$-$[Fe/H],
[Mg/Fe]$-$[Fe/H],
and [Mg/Ca]$-$[Fe/H]
relations between the eight dust wind LMC models 
(M2-M9) with different $\beta$ and $\gamma$ 
(for a fixed $C_{\rm r}$).
There is no substantial difference in the 
[Mg/Fe]$-$[Fe/H] relations between the dust wind models (M2-M5)
with different $\beta$
for a fixed $\gamma=-1$, because the dust removal fractions of Mg and Ca
are not so different for the ranges of these parameters, as shown in Fig. 1.
The [Ca/Fe]$-$[Fe/H]
and [Mg/Ca]$-$[Fe/H] relations depend on $\beta$ for a fixed $\gamma =-1$
in such a way that [Ca/Fe] and [Mg/Ca] can be systematically lower and higher,
respectively, at a given [Fe/H] for smaller $\beta$. The models with smaller 
$\beta$ are more consistent with observations among the four models
(M2-M5) with a fixed $\gamma$ in terms of reproducing the observed 
[Ca/Fe]$-$[Fe/H]
and [Mg/Ca]$-$[Fe/H] relations.
These results imply that if the observed
[Ca/Fe]$-$[Fe/H],
[Mg/Fe]$-$[Fe/H],
and [Mg/Ca]$-$[Fe/H] relations are all real,
then a steeper dependence of $P_i$ on $\delta_i$ is required for
reproducing these three relations in a self-consistent manner.

The dependences of the three relations on $\gamma$ for a fixed $\beta$ are not
so remarkable in comparison with those on $\beta$ (for $-1 \le \gamma \le -0.3$).
The models with smaller $\gamma$ show systematically lower [Ca/Fe] and higher
[Mg/Ca] for $\beta=0.3$ and 0.5, because the $f_{\rm dw, \it i}$ difference
between Ca, Mg, and  Fe is larger for smaller $\gamma$.  
The models with smaller $\gamma$ for a fixed $\beta$ can better reproduce
the observed
[Ca/Fe]$-$[Fe/H],
[Mg/Fe]$-$[Fe/H], and
[Mg/Ca]$-$[Fe/H] relations in a self-consistent manner.
These results confirm that $\gamma$ needs to be small so as to explain
the observed three relations reasonably well.
Owing to the observed larger dispersions of 
[Ca/Fe],
[Mg/Fe],
and [Mg/Ca]
at a given [Fe/H],
it is currently impossible for the present study to determine the best
set of $\beta$ and $\gamma$ for which the three observations can be 
most self-consistently explained.

\subsection{Prediction of lower [Ti/Fe]}

Since Ti is the second most severely dust-depleted element next to Ca among the observed
elements (e.g.,  Fig. 4 in SS96), the time evolution of Ti can be influenced
by dust wind to a larger extent than Mg and Fe in the present dust wind model.
It is therefore useful for the present study to provide some predictions
on the [Ti/Fe]$-$[Fe/H]
relation of the LMC by using the same LMC models discussed in preceding
sections. Fig. 5 shows the clear systematic differences in [Ti/Fe] 
evolution between
the LMC models with and without dust 
and (ii) the degrees of the [Ti/Fe] differences
depend on $\beta$ and $\gamma$.
However, the differences are not so large as those derived for [Ca/Fe] and [Mg/Ca],
which is expected for a smaller dust removal fraction for Ti.
The present study therefore predicts  that the LMC shows systematically lower [Ti/Fe] than
the MW and the degree of the [Ti/Fe] differences between the LMC and the MW
is smaller than that of Ca,
if the chemical evolution of the LMC is influenced by dust wind.
The results of P08  appear to be consistent with the above predictions,
though the observed dispersions of [Ca/Fe] and [Ti/Fe] at a given [Fe/H] are large.


\begin{figure}
\psfig{file=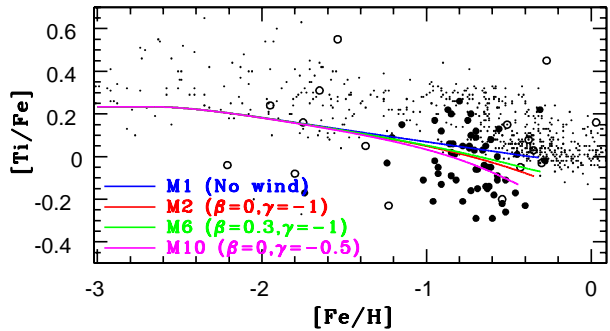,width=8cm}
\caption{
The evolution of [Ti/Fe]
as a function of [Fe/H]
for the eight LMC model with and without dust wind for different $\beta$ and $\gamma$.
Blue, red, green, and magenta lines
represent the models with
no dust wind (M1),
$\beta=0$ and $\gamma=-1$ (M2),
$\beta=0.3$ and $\gamma=-1$ (M6),
$\beta=0$ and $\gamma=-1$ (M10),
respectively.
}
\label{Figure. 5}
\end{figure}

\section{Discussion}

\subsection{The influences of dust sizes on the present results}

We have so far focused on the results of the models in which the possible differences
in sizes between different dust grains (in particular, Mg- and Ca-bearing ones)
are not included.
As discussed in \S 2.1.3,  the typical size
of Ca-bearing dust can be  possibly smaller than that of Mg-bearing one,
if Ca-bearing dust  can be preferentially formed  in SNe.
Furthermore, Kozasa \& Hasegawa (1987) showed that Ca-bearing dust (CaTiO$_3$) is typically
smaller than Mg-bearing one (MgSiO$_3$) in their theoretical models of dust formation in
cooling gas of the solar composition (e.g., see their Fig. 2).
F91 clearly showed that only
a factor of two difference in dust sizes (silicate vs graphite in their model,
see their Fig 7) can cause a significant difference in the time evolution
of the dust removal processes. Given that dust with smaller sizes can be more efficiently removed from
galaxies (F91),  Ca-bearing dust with its possibly smaller size would
be likely to be removed
even more efficiently than the present study derived.
Thus, the present results (which depend on more efficient removal of Ca than Mg)
can not be qualitatively changed by including the possible
size differences between Mg- and Ca-bearing dust, 
if Ca-bearing dust has a smaller size
than Mg-bearing one.

However, if the typical size of Mg-bearing dust is significantly
(by more than a factor
of two) smaller than that of Ca-bearing one, then the present results can be changed
qualitatively, because Mg-bearing dust can be more efficiently removed from
galaxies than Ca (even if the depletion level of Mg is lower than that of Ca).
We can not currently make a quantitative estimation
on the typical sizes of Mg- and Ca-bearing dust owing to the lack
of extensive observational and theoretical
works on the sizes of these dust.
Furthermore, 
there is no detailed theoretical work how $P_i$, which determines the dust
removal efficiency, depends on physical properties
of dust in ISM  such as dust sizes and compositions.
Therefore, it is reasonable for the present study to suggest that
the present results could be changed if the influences of the
typical sizes of Mg- can Ca-bearing on dust removal efficiencies
are properly modeled.
If future theoretical studies on the typical  sizes of Ca-bearing dust in
SNe and AGB stars provide a table for the sizes, we will be able to discuss
the influences of dust size on the present results
in a much more quantitative manner.

\begin{figure}
\psfig{file=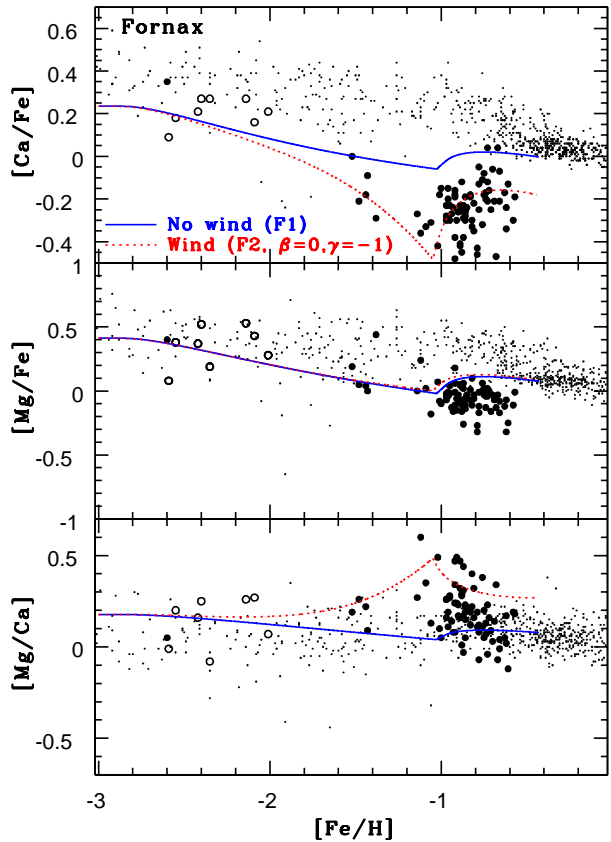,width=8cm}
\caption{
The evolution of [Ca/Fe] (top), [Mg/Fe] (middle), and [Mg/Ca] (bottom)
as a function of [Fe/H]
for the two Fornax models without dust wind (F1, blue solid)
and with dust wind (F2, red dotted).
In these Fornax models, a secondary starburst with a high SFR at $t=5$ Gyr is assumed.
The observational results shown for the MW are exactly the same as those in Fig. 2.
The results for clusters (big open circles) and field stars (big filled circles)
in the Fornax dwarf galaxy
from Letarte et al. (2010) are plotted in this figure.
}
\label{Figure. 6}
\end{figure}

\begin{figure}
\psfig{file=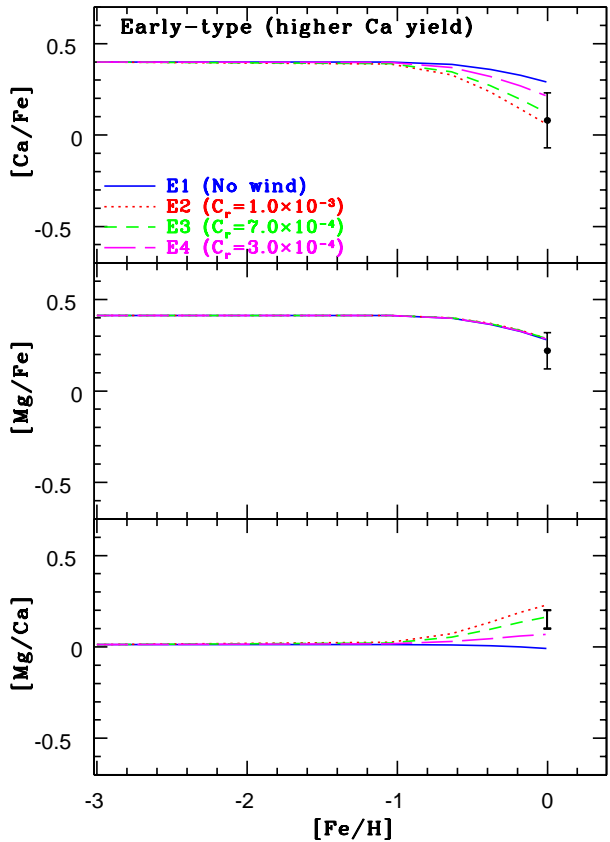,width=8cm}
\caption{
The evolution of [Ca/Fe] (top), [Mg/Fe] (middle), and [Mg/Ca] (bottom)
as a function of [Fe/H]
for the four early-type  models
with and without dust wind for different $C_{\rm p}$ (but fixed $\beta=0$ and $\gamma=-1$).
Blue, red, green, and magenta lines
represent the models with
no dust wind (E1),
$C_{\rm r}=10^{-3}$  (E2),
$C_{\rm r}=7 \times 10^{-4}$  (E3),
and $C_{\rm r}=3 \times 10^{-4}$  (E4),
respectively.
The observational data for mean values of [Ca/Fe] and [Mg/Fe]
(for [Fe/H]$=0$)  from J12 are plotted with an observational error bar (0.1 and 0.15 dex)
in the top and middle panels. The observed range of [Mg/Ca] is shown in the bottom panel.
For these early-type models, higher Ca yield is used so that both initial
[Mg/Fe] and [Ca/Fe]
can be as high as those ($\sim 0.4$) of the old Galactic halo stars.
}
\label{Figure. 7}
\end{figure}

\subsection{Are there any other explanations  for high [Mg/Ca] ?}

The present study has shown, for the first time, that both the observed rather high [Mg/Ca]
($>0.3$) and the increasing trend of [Mg/Ca] 
with increasing [Fe/H] (at [Fe/H]$>-1$) in the LMC can be self-consistently
reproduced by the dust wind model for a reasonable set of the model parameters.
Although this can be regarded as remarkable improvement of our chemical evolution model
over BT12,  it should be noted that there could be alternative explanations
for the observed high [Mg/Ca]. Since different IMFs and selective SN winds
can not self-consistently explain the observed [Mg/Fe]$-$[Fe/H] and [Ca/Fe]$-$[Fe/H]
relations, we consider the following two alternative explanations to the dust wind one.

One is that P08 could  overestimate the [Mg/Ca]
of stars in the LMC for some reasons. This seems to be the case, given that [Ca/Fe] and [Mg/Fe]
derived by V13 for the LMC inner disk and bar are higher and lower, respectively, than
those by P08. The reason for this difference between P08 and V13 is 
that they  adopted  different
stellar parameters and methods to derive stellar abundances. If the abundance determination
by V13 is more accurate, then the observed [Mg/Ca] in P08 can be lowered by $\sim 0.3$ dex
and thus become closer to the solar value for
most of the investigated stars in the LMC. However, this possible 0.3 dex reduction
of [Mg/Ca] is not enough to explain the observed high [Mg/Ca] ($>0.4$) of some stars
in P08.

The other is that high [Mg/Ca] is due to the contribution of jet-induced SNe that have
characteristic nucleosynthesis yields.
Using chemodynamical models with explosions of aspherical SNe (A-SNe),
Bekki et al. (2013) investigated how the characteristic chemical yields of A-SNe can influence
the chemical evolution of dwarf galaxies. They found that [Mg/Ca] can be quite diverse in
different stars for a given [Fe/H] with some stars having  as large as [Mg/Ca]$\sim 0.5$:
The physical reason for this is discussed in their paper.
Although their results imply that the observed rather high [Mg/Ca] in the LMC can be due to
the contribution of A-SNe to the chemical evolution,  they could 
not explain  the increasing trend
of [Mg/Ca] with increasing [Fe/H] in their models. 
Therefore, the A-SN scenario can not explain both the observed high [Mg/Ca]
and [Mg/Ca]$-$[Fe/H] relation of the LMC.

Thus, the dust wind scenario appears to be more promising than the 
above A-SNe scenario.
The dust wind scenario predicts that dust abundance patterns and compositions 
in outer gaseous
halos of galaxies, where interstellar dust can reach owing to radiation pressure of stars
under some physical conditions (F91),  should be different from those in the main bodies
of the host galaxies. For example, the Ca-abundance of dust in gaseous halos could be significantly
higher than that in ISM of disks owing to the differential removal process of dust.
Therefore, 
such  predictions need to be investigated by observational studies of dust abundances 
and compositions of gaseous halos of galaxies so that the viability of the scenario can be 
assessed in a convincing way.
Recent observational studies of extinction curves for different halo regions of the Galaxy
(e.g., Peek et al. 2013)
would provide some information on how the dust abundances and compositions  of the halo
are different from those of the disk.

\subsection{Differential dust removal in other galaxies ?}

\subsubsection{The MW}

As shown in Figs. 2 and 3,  the MW does not show low [Ca/Fe] and high [Mg/Ca]
for the same metallicity range of the LMC stars. This striking contrast in the
[Ca/Fe]$-$[Fe/H] and [Mg/Ca]-[Fe/H] relations between the two galaxies can be understood
in the context of the different dust removal efficiencies between the two
as follows. Although dust can be possibly removed once from the main disk of the MW through
dust wind,  a large fraction  of the dust can be return back to the original disk
to participate the further chemical evolution owing to the deep gravitational potential.
On the other hand, the LMC dust can not be returned back to its original disk
once it is removed through dust wind, because the shallow gravitational potential
can not keep it within the inner halo of the LMC.
The strong tidal field of the Galaxy could also prevent the once removed dust
from returning back to its original disk of the LMC.
Thus, the dust removal efficiency can be dramatically different
between the LMC and the MW in the sense that  the efficiency
is much lower in the MW: the chemical evolution of the MW
is less likely to be  strongly influenced by dust wind.
In order to discuss 
whether this explanation is reasonable and realistic,
we will perform numerical simulations of dust wind for galaxies with different
gravitational potentials in our future works.

\subsubsection{Fornax dwarf galaxy}
Recent spectroscopic observations of stars in the Fornax dwarf galaxy (e.g., Letarte et al. 2010)
have revealed that [Ca/Fe] at [Fe/H]$\ge -1$ is  significantly smaller than the solar value while 
[Mg/Fe] is consistent with the solar value. These trends in the Fornax galaxy are similar to those
found for the LMC and summarized  in Fig. 6.  One of possibly significant
differences in the chemical
abundance pattern between the LMC and the Fornax galaxy is that 
[Ca/Fe] decreases with [Fe/H] until [Fe/H]$\sim -1$ and then it appears to start to 
increase from [Fe/H]$\sim -1$. 
Below, we briefly discuss whether the dust wind model can explain the observed abundance patterns
of the Fornax galaxy.  A fully self-consistent chemical evolution model with dust wind
for the Fornax galaxy should be 
constructed in our future paper.

Both recent observational and theoretical studies have shown
that the Fornax galaxy could have experienced secondary starbursts possibly triggered by
merging of two or multiple dwarf galaxies
(e.g., Coleman et al. 2004; Tsujimoto 2011; Yozin \& Bekki 2012;
de Boer et al. 2013).
Some of recent observational studies have suggested that the starburst 
events in the Fornax dwarf galaxy can be once between 6 and 9 Gyr ago
(e.g., Piatti et al. 2014) and several times in the early formation
histories of the galaxy (e.g., Hendricks et al. 2014). Therefore,
it is not reasonable for this study to adopt a uniform star formation
history without secondary starburst events.

Thus, we consider a 
one-zone chemical evolution model with a secondary starburst. 
In this model, the star formation coefficient ($C_{\rm sf}$) can become rather high
during the starburst period. We assume that (i) $C_{\rm sf}$ is 0.003 before the starburst
and 0.04 after the starburst, (ii) the starburst occurs during $t=5$ and 6 Gyr,
(iii) SF is truncated after the starburst,
and (iv) $f_{\rm b}$, $\beta$, $\gamma$, and 
$C_{\rm r}$ are set to be 0.09, 0, $-1$,  and $6 \times 10^{-3}$, respectively,
 in this model.
These values of basic model parameters are chosen such that
both the observed time evolution of [Ca/Fe] and [Mg/Fe]
can be better reproduced.
Like the LMC model,  these Fornax dwarf models do not include
stellar wind from SN explosion for clarity.
For comparison, a model with no dust wind yet
the same parameter values as those described
above is investigated.

Fig. 6 shows that [Ca/Fe] can become rather low (up to $-0.5$) at [Fe/H]$<-1$ before the
secondary starburst owing to the stronger dust wind assumed in this model. 
The increasing trend of [Mg/Ca] with increasing [Fe/H] can be clearly seen in the dust wind model,
and [Mg/Ca] can be as large as 0.5 at [Fe/H]$\sim -1$. 
Both [Ca/Fe] and [Mg/Fe] can rapidly increase with the [Ca/Fe] evolution
being more remarkable
during the secondary
starburst so that [Mg/Ca] can finally become smaller ($\sim 0.2$).
The modeled [Mg/Fe] after the starburst is slightly higher than the observed one, which implies
that this dust wind model is less consistent with observational results for the Fornax galaxy.
However, the wind model implies that radiation-driven dust wind can be responsible
for the observed rather low [Ca/Fe] and high [Mg/Ca].
It is interesting to point out that [Ti/Fe] can be slightly larger than [Ca/Fe] for 
the Fornax galaxy
(See Fig. 10 in Letarte et al. 2010) and this trend is consistent with the prediction of
the dust wind model.
It is our future study to construct a better model to explain both the observed [Mg/Fe] and [Ca/Fe]
in the Fornax galaxy.

In the above discussion, it is assumed that the dust removal process
is essentially the same between the LMC and the Fornax galaxy, which 
could be a
remnant of dwarf-dwarf merging (i.e., the same model for $P_i$ dependent
only on dust depletion levels). It would be possible that dust 
can be efficiently removed from merging galaxies owing to tidal
stripping. Since tidal stripping process does not depend on chemical elements
of ISM and phases of ISM (cold or hot), 
it is unlikely that only a specific type of dust (e.g., Ca-bearing dust)
can be preferentially stripped: All of elements would be stripped
equally through tidal stripping during merging.
 Therefore, galaxy merging might not result
in differential dust removal required for explaining the observed
chemical abundances of the Fornax galaxy.

\subsubsection{Early-type galaxies}

Thomas et al. (2003, T03) investigated the abundances of various  $\alpha$ elements in 
early-type galaxies and found that Ca abundance is smaller with respect to the other $\alpha$ 
elements by a factor of $\sim 2$. They also found that [$\alpha$/Ca] is larger
for early-type galaxies with larger velocity dispersion. Indeed, some of early-type galaxies
in Fig. 2 of T03 shows [$\alpha$/Ca] as large as and larger than 0.1.
These  results imply that [Mg/Ca] of some early-type galaxies can be 
higher than the solar value, though the level of [Mg/Ca] enhancement in the galaxies
is not so high as that of the LMC. The latest observational results 
by Johansson et al. (2012, J12)
are essentially the same as those in T03, which suggests that the [Mg/Fe]
($0.1 \sim 0.2$ dex) higher than [Ca/Fe]  in T03
is real. J12 derived the relations between the velocity dispersions ($\sigma$ km s$^{-1}$)
and the chemical abundances (e.g., [Fe/H], [Mg/Fe], and [Ca/Fe]) of elliptical
galaxies. These can be used for discussing the origin of [Ca/Fe] and [Mg/Ca]
of elliptical galaxies in the present study.

Here we briefly discuss the origin of the higher [Mg/Ca] in early-type galaxies by using 
the dust wind model in which model parameters can be  reasonable for early-type galaxies.
In this dust wind model for early-type galaxy formation,
$t_{\rm a}$, $t_{\rm end}$, $C_{\rm sf}$, $\beta$, and $\gamma$  are set to be
0.1 Gyr, 0.5 Gyr, 0.6, 0, and $-1$, respectively. 
These values of model parameters are chosen such that
(i) elliptical galaxies  can experience initial massive starbursts
and (ii) both  final [Fe/H] and [$\alpha$/Fe] can be high.
The stellar wind from SN explosion is not included in these elliptical
galaxy models.
In these models, 
Ca yield
($Y_{\rm Ca}$) is higher than the
theoretical yield of Ca from T95 ($Y_{\rm Ca,0}$)
so that both the initial [Mg/Fe] and [Ca/Fe] can be as high as those of the
Galactic old halo stars.
The parameter $C_{\rm r}$ is changed
so that we can see the dependence of galactic chemical evolution of the strength of
radiation pressure on dust grains. For comparison, a model in which dust wind is not
included and model parameters are exactly the same as those in the dust wind model
is investigated. These models can show the final [Fe/H] and [Mg/Fe] as large as 0 and 
and 0.2, respectively, which are reasonable for luminous early-type galaxies.
Based on the observational results in J12 (their Figs. 9, 11, and 13),
we can derive the mean [Mg/Fe], [Ca/Fe],
and [Mg/Ca] for [Fe/H]$=0$ (i.e.,  $\log \sigma =2.2$).

Fig. 7 shows that if $C_{\rm r}$ is as large as $7 \times 10^{-4}$ (i.e., 70\% 
of  the one adopted for the LMC model), then the final [Mg/Ca] can be
consistent with the observed range of [Mg/Ca].
Owing to the rather high SFRs thus high radiation pressure
on dust in these dust wind models,
a significant fraction of Ca can be removed from the main galaxies. 
The model with no dust wind, on the other hand, shows a lower final
[Mg/Ca] ($\sim 0$). These results imply that the observed higher [Mg/Ca] of
some bright elliptical galaxies could be 
due to the differential dust removal driven by radiation pressure of stars in forming
early-type galaxies. 
The observed mean [Mg/Ca] at [Fe/H]=0 in J12 can be consistent with the model E3
whereas the observed mean [Ca/Fe] is best fit to the model E2-E4 with dust wind.
It should be stressed here that Fig. 7 just illustrates one possible scenario for the observed
higher [Mg/Ca] in early-type galaxies. There could be other explanations
for the observed underabundance of [Ca/Fe] and higher [Mg/Ca], such as 
metallicity-dependent SN yields (T03), which needs to be explored by our future studies.
Thus, the present study suggests that the observed higher [Mg/Ca] in massive dwarfs and early-type
galaxies can be explained in terms of radiation-driven dust wind in galaxies.

\section{Conclusion}

We have investigated how radiation-driven dust wind can influence galactic chemical evolution
by using our new one-zone chemical evolution models. We have particularly focused on the time evolution
of [Fe/H], [Mg/Fe], [Ca/Fe], and [Mg/Ca] in massive dwarf galaxies (e.g., LMC),
because these galaxies are observed to have intriguing abundance patterns that were not explained
reasonably well by our previous models without dust wind. By comparing between the present
new results and the latest observations, we have tried to find how the model parameters for dust
wind can control the time evolution of the correlations between
[Fe/H], [Mg/Fe], [Ca/Fe], and [Mg/Fe].
Although the adopted dust wind model is somewhat idealized, we have found the 
following preliminary results. \\

(1) The time evolution of [Ca/Fe] can be significantly different between the models with
and without dust wind in the sense that [Ca/Fe] at a given [Fe/H] is lower in the dust wind model.
On the other hand, [Mg/Fe] at a given [Fe/H] is not different between the two models with and without
dust wind. This is mainly because Ca is more severely dust-depleted than Mg and Fe
so that Ca can be more efficiently removed from galaxies through radiation-driven dust wind.
This `differential dust removal'  is the main physical mechanism for the derived different evolution
of [Ca/Fe]  and [Mg/Fe]  in the present models. \\

(2) As a result of differential dust removal,  the time evolution of [Mg/Ca] can be significantly
different between the models with and without dust wind.  The higher [Mg/Ca] in the dust wind model
is more consistent with the observed [Mg/Ca] in the LMC and Fornax. Furthermore,
the dust wind model predicts an increasing trend of [Mg/Ca] with [Fe/H], which also appears to be
consistent with observations for these massive dwarfs. It should be noted, however,
that the observed [Mg/Ca] show a large scatter and the latest observations of [Mg/Ca] have 
lowered  [Mg/Ca] (V13) in  comparison with previous observations (P08). \\

(3) [Ti/Fe]$-$[Fe/H] relation can be significantly different between the models with and without
dust wind. However, the difference is not so large in comparison with the  [Ca/Fe]$-$[Fe/H] relation,
because Ti is the second most severely dust-depleted next to Ca
among the observed  elements.
These predicted trends are apparently observed in the inner disk of the LMC and Fornax. \\

(4) Final [Ca/Fe] and [Mg/Ca] in the early-type  galaxy models
with dust wind can be lower and higher, respectively, than the models without dust 
wind. This is mainly because much stronger radiation-driven wind in the initial starburst
phases of elliptical galaxy formation can remove their dust quite efficiently. The derived
lower [Ca/Fe] and higher [Mg/Ca] can provide a new clue to the origin of the observed
[Ca/Fe] and [Mg/Ca] in elliptical galaxies. \\

(5) The differential dust removal process depends basically on the two parameters,
$\beta$ and  $\gamma$
(which controls the minimum level of the dust removal efficiency  and the dependence
of the efficiency on the dust depletion level, respectively).  For a given $\beta$,
differences in [Ca/Fe] between models with and without dust wind
can be larger for smaller $\gamma$ (i.e., steeper dependence of dust-removal efficiency
on dust-depletion level). 
For a given $\gamma$,
the [Ca/Fe] differences between the two models
can be larger for smaller $\beta$.
Although the adopted functional form of $P_{\rm i}$ 
(depending on $\beta$ and $\gamma$) can be reasonable, 
these two parameters are currently very hard to be constrained by observations.  \\

(6) These results in (1)-(5) 
 suggest that galactic chemical evolution can be influenced strongly
by galactic luminosity evolution, because the radiation-driven dust removal processes in
a galaxy,
which depend primarily on the luminosity  evolution of the galaxy,
can reduce the total amount of gas-phase metals in the main body of the galaxy.
The results also suggest that we would need to understand how the physical processes
of gas-phase metals being locked up in dust grains depend on the compositions and sizes of dust 
for each individual element in order to model the differential dust removal processes
in a more sophisticated way.
This is because the time evolution of dust grains under radiation pressure of stars
depends on dust compositions and sizes (F91). \\

(7) If the differential dust removal modeled in the present study is correct,
then we predict that
the outer gaseous halos of galaxies, where dust wind can reach under some physical conditions,
should have different dust abundances and compositions (e.g., higher Ca abundance) in comparison
with ISM of disks. Therefore, future observational studies of dust abundances and compositions
in the outer halos of galaxies (far beyond the optical disks)
will provide a stringent test for the dust wind model proposed in the present study. \\

\section{Acknowledgment}
We are   grateful to the referee, Mathieu van der Swaelmen,  for his
constructive and useful comments that improved this paper.

\appendix

\section{Dependences on Ca yield}

\begin{figure}
\psfig{file=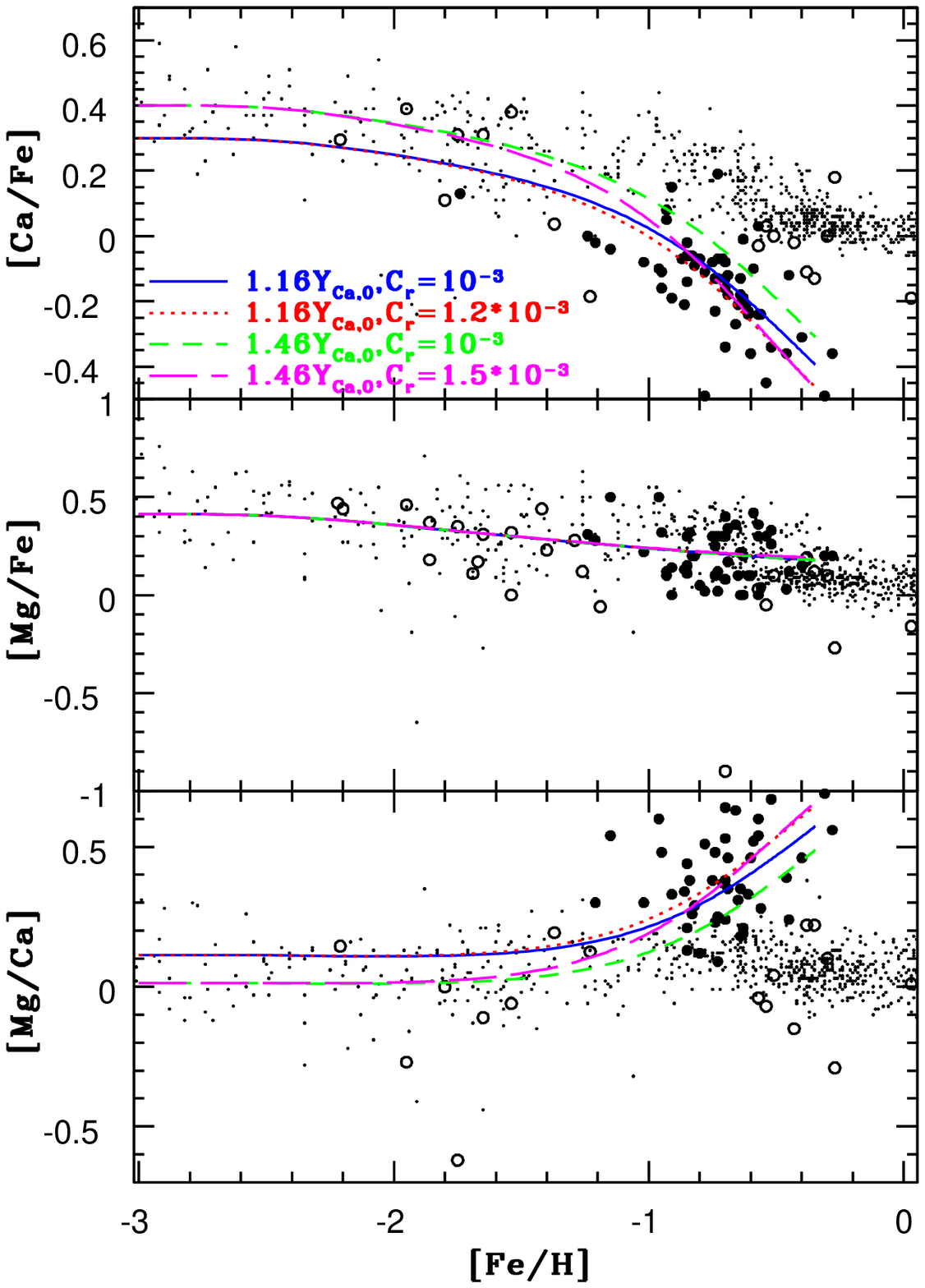,width=8cm}
\caption{
The same as Fig. 4 but for only four models with different Ca yield
($Y_{\rm Ca}$) and $C_{\rm r}$.
The theoretical yield of Ca from T95 is referred to as $Y_{\rm Ca,0}$
for convenience.
The blue, red, green, and magenta lines
represent the models with
$Y_{\rm Ca}=1.16Y_{\rm Ca, 0}$ and $C_{\rm r}=10^{-3}$,
$Y_{\rm Ca}=1.16Y_{\rm Ca, 0}$ and $C_{\rm r}=1.2 \times 10^{-3}$,
$Y_{\rm Ca}=1.46Y_{\rm Ca, 0}$ and $C_{\rm r}=10^{-3}$,
and $Y_{\rm Ca}=1.46Y_{\rm Ca, 0}$ and $C_{\rm r}=1.5 \times 10^{-3}$,
respectively.
For higher Ca yields, more efficient dust removal (corresponding to
larger $C_{\rm r}$)  is required for explaining the observed very low [Ca/Fe].
}
\label{Figure. A1}
\end{figure}

BT12 and the present study used the same {\it theoretical}
chemical yields shown in T95 for consistency
between the two works on the LMC chemical evolution. The predicted [Ca/Fe]
at lower metallicity ([Fe/H]$<-2$) in the present LMC models
can be  therefore $\sim 0.24$.
The [Ca/Fe] in the early phases of the LMC is not so high as $0.3 \sim 0.4$ observed
in the metal-poor halo stars of the Galaxy. 
One might have a concern  that the initially lower [Ca/Fe] could 
be partly responsible for the present successful LMC models reproducing the observed
very low [Ca/Fe]. In order to remove this concern, we have investigated
the models with higher Ca yields for which the initial [Ca/Fe] can be as high
as $0.3 \sim 0.4$. 

Fig. A1 shows that for the model with $Y_{\rm ca}=1.16 Y_{\rm Ca,0}$,
the observed low [Ca/Fe] can be well reproduced by adopting
slightly higher $C_{\rm r}$ corresponding to a  more efficient
dust removal process from the main body of the LMC.
This result can be seen in the model with
$Y_{\rm ca}=1.46 Y_{\rm Ca,0}$,
which means that the essential influences of the dust wind on the LMC chemical
evolution do not depend on the adopted Ca yield.
These results therefore  strengthen the present most important
conclusion that the dust wind can be  primarily responsible for the origin of
low [Ca/Fe] and high [Mg/Ca] in the LMC.

\section{The case of a steeper IMF}

Fig. B1 shows how dust wind can influence the chemical evolution of the LMC
in the models with a steeper IMF ($\alpha=2.55$). In these models,
the value of $C_{\rm sf}$ is set to be 0.01 (instead of 0.006 adopted
for $\alpha=2.35$), because final [Fe/H] can be as high as $-0.3$ 
owing to the higher SFRs and chemical enrichment processes for such
an adoption of $C_{\rm sf}$. Clearly, the models with dust wind can 
much better reproduce the observed chemical abundances than the model with no dust wind.
This strongly suggests  that the roles of dust wind in galactic chemical evolution
do not depend on the adopted IMFs.
Furthermore, these steeper IMF models with dust wind show systematically lower
[Ca/Fe] for a given set of $\beta$ and $\gamma$. 
This results is not so surprising, because our previous models (BT12) have already
shown that IMF slopes can control the time evolution of [Ca/Fe].
The dust wind model with $\beta=0.3$ and $\gamma=-1$ (i.e., smaller difference
in dust removal efficiency between Ca and Mg owing to the larger $\beta$) can
show the highest [Ca/Fe] and the lowest [Mg/Ca] in these models with dust wind.

\begin{figure}
\psfig{file=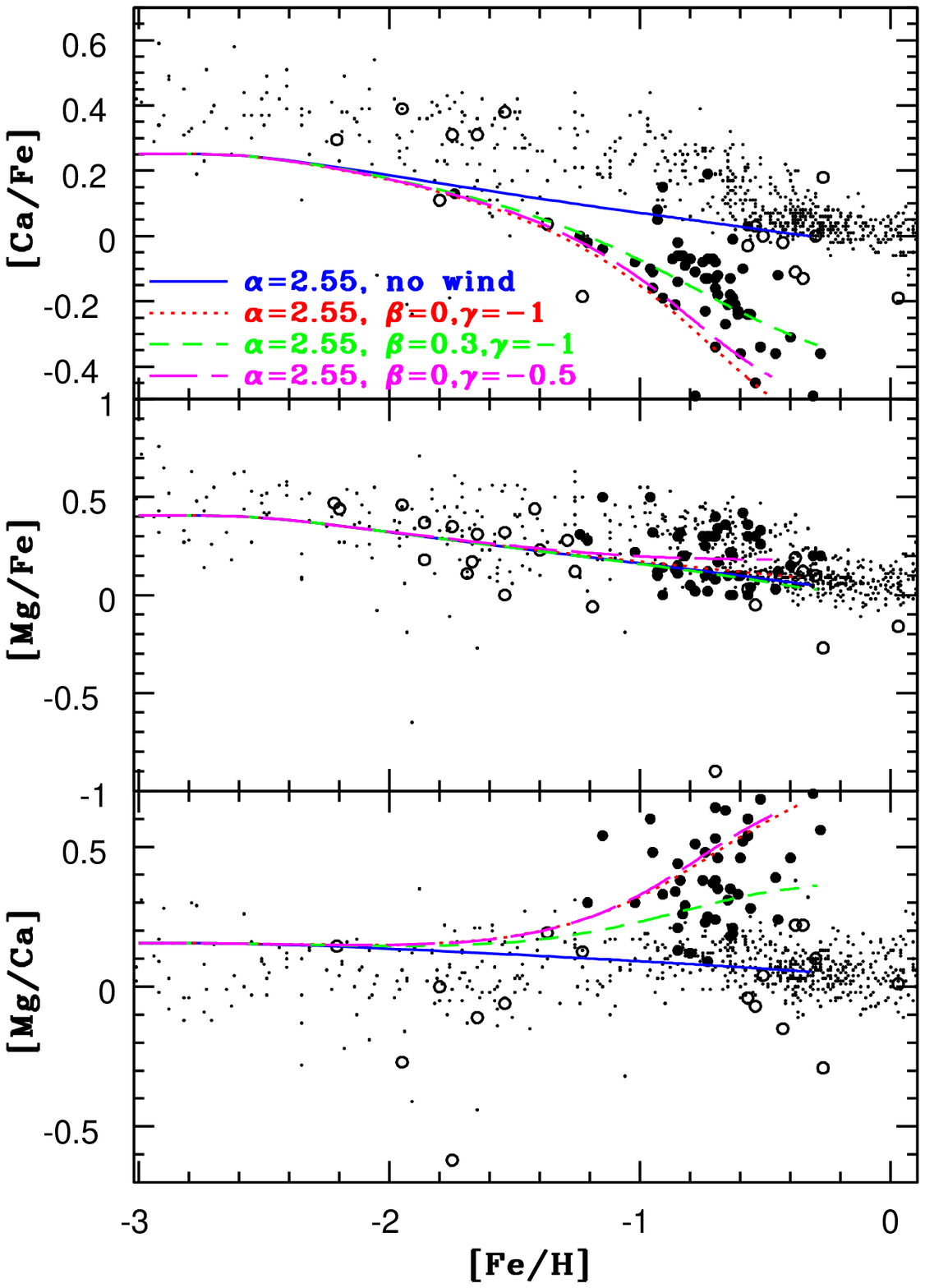,width=8cm}
\caption{
The same as Fig. 4 but for only four models with a same steeper IMF
($\alpha=2.55$) yet different $\beta$ and $\gamma$.
The blue, red, green, and magenta lines
represent the models with
no dust wind,
$\beta=0$ and $\gamma=-1$,
$\beta=0.3$ and $\gamma=-1$,
and $\beta=0$ and $\gamma=-0.5$,
respectively.
}
\label{Figure. B1}
\end{figure}

\section{The roles of stellar winds in chemical evolution}

Fig. C1 show the time evolution of chemical abundances of the LMC
in the models with stellar wind ($\alpha=2.35$). As assumed in BT12,
40\% of ejecta from SNII and SNIa is removed completely from ISM of the LMC
(i.e., never returned back to original ISM) in these models. This is
a `selective wind' model in the sense that only SN ejecta can be removed
from ISM (BT12).
The value of $C_{\rm sf}$ is set to be 0.01 (instead of 0.006 adopted
for the model with no stellar wind and $\alpha=2.35$) in these models. This is  because 
final [Fe/H] can not  be as high as $-0.3$ 
owing to the loss of metals through SN wind,
if $C_{\rm sf}=0.006$ is assumed.
The adopted larger $C_{\rm sf}$ results in higher SFRs and thus 
more efficient chemical 
enrichment processes so that the final [Fe/H] can be similar to the observed
value.

The roles of dust wind in the chemical evolution of the LMC
can be clearly seen in these four models with stellar wind. This result
can strengthen the present key conclusion that more efficient removal
of Ca (in comparison of Mg) can be responsible for the observed low
[Ca/Fe] and high [Mg/Ca] in the LMC. The dependences of the chemical evolution
of [Ca/Fe], [Mg/Fe], and [Mg/Ca] on $\beta$ and $\gamma$ are essentially
the same as those derived in other models. The model with a steeper dependence
of $P_i$ on dust depletion levels ($\beta=0$ and $\gamma=-1$)
shows rather low [Ca/Fe] and high [Mg/Ca] in these wind models.

\begin{figure}
\psfig{file=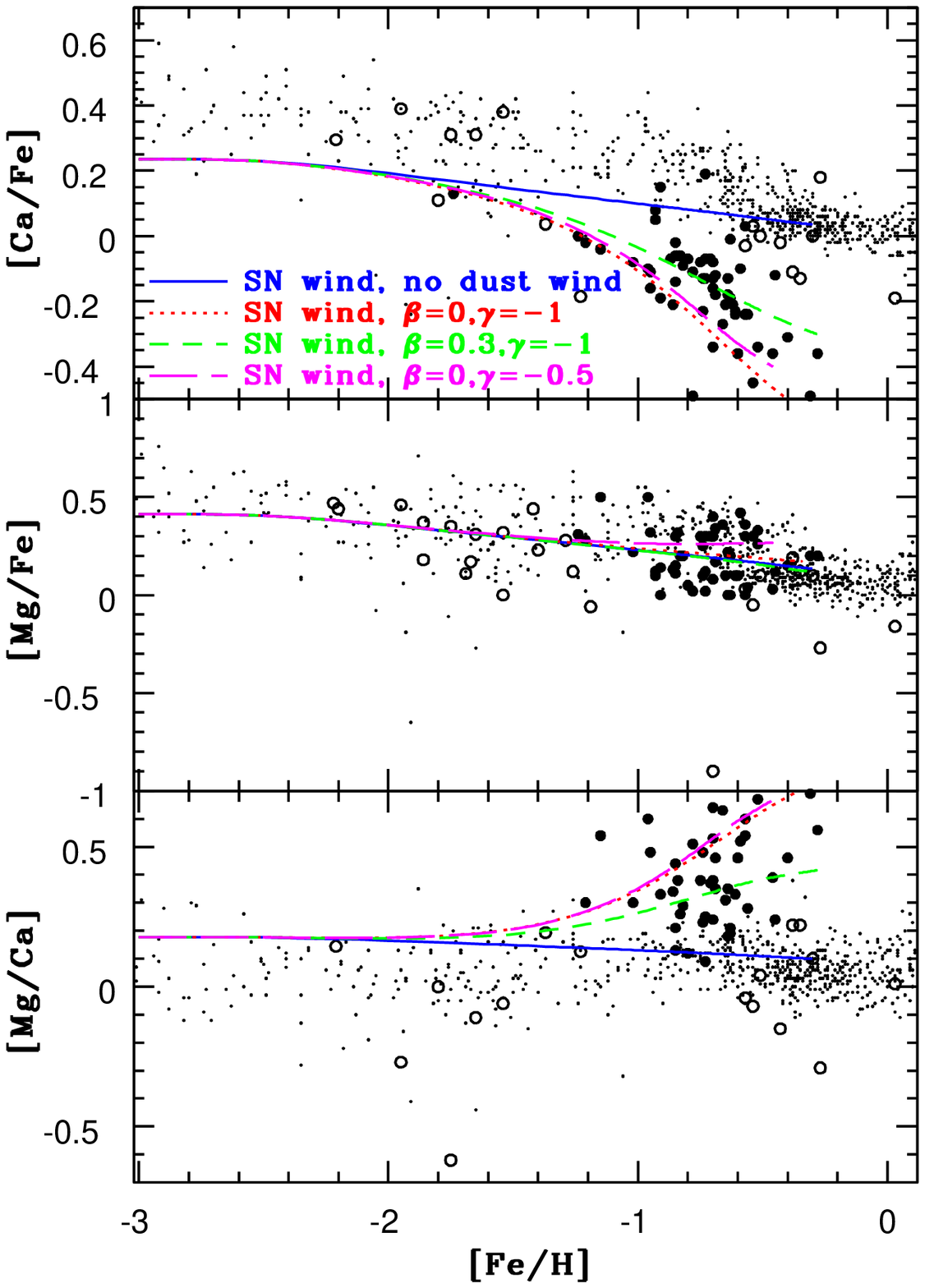,width=8cm}
\caption{
The same as Fig. 4 but for only four models with the Salpeter IMF,
stellar wind from SN explosions ('SN wind'),
and different  $\beta$ and $\gamma$.
The blue, red, green, and magenta lines
represent the models with
no dust wind,
$\beta=0$ and $\gamma=-1$,
$\beta=0.3$ and $\gamma=-1$,
and $\beta=0$ and $\gamma=-0.5$,
respectively.
In these models with stellar wind,
the 'selective wind model' adopted by BT12, in which only SN ejecta
can be removed from  the LMC through wind,  is adopted.
}
\label{Figure. C1}
\end{figure}


\begin{thebibliography}{}

\bibitem[]{}
Aguirre, A., Hernquist, L., Katz, N., Gardner, J.,   Weinberg, D.,
2001, ApJ, 556, L11

\bibitem[]{}
Barsella, B., Ferrini, F., Greenberg, J. M.,   Aiello, S.,
1989, A\&A, 209, 349

\bibitem[)]{}
Bekki, K., 2013, MNRAS, 432, 2298 

\bibitem[]{}
Bekki, K., 2014, submitted to MNRAS

\bibitem[]{}
Bekki, K.,  Tsujimoto, T., 2012, ApJ, 761, 180 (BT12)

\bibitem[)]{}
Bekki, K., Shigeyama, T.,   Tsujimoto, T., 2013, MNRAS, 428, L31



\bibitem[]{}
Chiao, R. Y.,   Wickramasinghe, N. C., 1972, MNRAS, 159, 361

\bibitem[]{}
Coker, C. T., Thompson, T. A.,  Martini, P.,
2013, ApJ, 778, 79

\bibitem[]{}
Coleman, M. G., Da Costa, G. S.,  Bland-Hawthorn, J.,  et al. 2004, AJ, 127, 832

\bibitem[Colucci et al.(2012)]{Colucci_12}
Colucci, J. E., Bernstein, R. A., Cameron, S. A.,  McWilliam, A.,
 2012, ApJ, 746, 29 (C12) 

\bibitem[)]{}
Davies, J. I., Alton, P., Bianchi, S.,   Trewhella, M.,
1998, MNRAS, 300, 1006

\bibitem[]{}
de Boer, T. J. L., Tolstoy, E., Saha, A.,  Olszewski, E. W., 2013, A\&A, 551, 103

\bibitem[)]{}
Draine, B. T., 2009, Physics of the interstellar and intergalactic medium

\bibitem[]{}
Dwek, E. 1998, ApJ, 501, 643 

\bibitem[)]{}
Ferrara, A., Ferrini, F., Barsella, B., \& Franco, J.
1991, ApJ, 381, 137 (F91)

\bibitem[]{}
Ferrarotti, A. S., \&  Gail, H.-P., 2005, A\&A, 430, 959

\bibitem[]{}
Ferrarotti, A. S., \&  Gail, H.-P., 2006, A\&A, 447, 553

\bibitem[)]{}
Franco, J., Ferrini, F., Barsella, B., \&  Ferrara, A.,
1991, ApJ, 366, 443



\bibitem[]{} 
Hendricks, B., et al., 2014, ApJ, 785, 102


\bibitem[]{} 
Holwerda, B. W., Keel, W. C., Williams, B., Dalcanton, J. J., \&  de Jong, R. S.
2009, AJ, 137, 3000

\bibitem[]{} 
Haschke, R., Grebel, E. K., \&  Duffau, S. 2012, AJ, 144, 106

\bibitem[]{}
Hirashita, H., 1999, ApJ, 522, 220

\bibitem[]{}
Hirashita, H., 2013, in the conference procceding, "The life cycle of dust in the Universe"

\bibitem[]{}
Jenkins, E. B. 2009, ApJ, 700, 1299

\bibitem[]{}
Johansson, J., Thomas, D., \& Maraston, C. 2012, MNRAS, 421, 1908 (J12)

\bibitem[Johnson et al.(2006)]{Johnson_06}
Johnson, J. A., Ivans, I. I., \& Stetson, P. B. 2006, ApJ, 640, 801

\bibitem[]{} 
Kirby, E. N., Cohen, J. G., Smith, G. H., Majewski, S. R.,
Sohn, S. T.,   Guhathakurta, P., 2011, ApJ, 727, 79

\bibitem[]{}
Kozasa, T., Hasegawa, H., 1987, PThPh, 77, 1402

 
\bibitem[]{}
Lanfranchi, G. A., \&  Matteucci, F. 2012, A\&A, 512, 85

\bibitem[]{}
Letarte, B., et al. 2010, A\&A, 523, 17

\bibitem[Maoz et al.(2010)]{Maoz_10}
Maoz, D., Sharon, K., \& Gal-Yam, A. 2010, ApJ, 722, 1879

\bibitem[]{}
Maoz, D., Mannucci, F.,  Li, W.,  Filippenko, A. V.,
Della Valle, M.,  \&  Panagia, N.
2011, MNRAS, 412, 1508

\bibitem[]{}
Maoz, D., \& Badenes, C. 2010, MNRAS, 407, 1314

\bibitem[]{}
Matteucci, F., \&  Francois, P. 1989, MNRAS, 239 885

\bibitem[]{}
McKee, C. F., \&  Ostriker, J. P., 1977, ApJ, 218, 148

\bibitem[]{}
McWilliam, A., Wallerstein, G., \&  Mottini, M. 2013, ApJ, 778, 149

\bibitem[]{}
M\'enard, B.,  Scranton, R., Fukugita, M., \&  Richards, G. 2010, MNRAS, 405, 1025

\bibitem[Mucciarelli et al.(2008)]{Mucciarelli_08}
Mucciarelli, A., Carretta, E., Origlia, L., \& Ferraro, F. R. 
2008, AJ, 136, 375 

\bibitem[Mucciarelli et al.(2010)]{Mucciarelli_10}
Mucciarelli, A., Origlia, L., \& Ferraro, F. R. 2010, ApJ, 717, 277 (M10)

\bibitem[Mucciarelli et al.(2011)]{Mucciarelli_11}
Mucciarelli, A., et al. 2011, MNRAS, 413, 837 (M11)

\bibitem[]{}
Murray, N., Quataert, E., \&  Thompson, T. A.,
2005, ApJ, 618, 569

\bibitem[]{}
Nozawa, T., Kozasa, T.,  Umeda, H.,  Maeda, K.,  \& Nomoto,  K., 2003, ApJ, 598, 785


\bibitem[Pagel \& Tautvai\v{s}ien\'{e}(1998)]{Pagel_98}
Pagel, B. E. J., \& Tautvai\v{s}ien\'{e}, G. 1998, MNRAS, 299, 535 


\bibitem[]{}
Peek, J. E. G., \&  Schiminovich, D. 2013, ApJ, 771, 68

\bibitem[]{}
Piatti, A. E.,  del Pino, A.,  Aparicio, A.,  Hidalgo, S. L.,
2014, MNRAS in press (arXiv1406.5911)

\bibitem[]{}
Piovan, L., Chiosi, C., Merlin, E., Grassi, T., Tantalo, R.,
Buonomo, U., Cassara, L. P., 2011, in preprint (arXiv1107.4541)


\bibitem[]{}
Pipino, A., Fan, X. L., Matteucci, F., Calura, F., Silva, L., Granato, G., \&  Maiolino, R.
2011, A\&A, 525, 61

\bibitem[Pomp\'{e}ia et al.(2008)]{Pompeia_08}
Pomp\'{e}ia, L., et al. 2008, A\&A, 480, 379 (P08)

\bibitem[]{}
Recci, S., Hensler, G., 2013, A\&A, 551, 41

\bibitem[]{}
Roedere, I. U., \& Kirby, E. N. in preprint (arXiv:1403.2733)


\bibitem[]{}
Roussel, H., et al. 2010, A\&A, 518, L66

\bibitem[]{}
Ruiz, L. O., Falceta-Goncalves, D., Lanfranchi, G. A., Caproni, A.,
2013, MNRAS, 429, 1437

\bibitem[]{}
Salpeter, E. E. 1955, ApJ, 121, 161

\bibitem[]{}
Savage, B. D., \& Sembach, K. R.  1996, ARA\&A 34, 279 (SS96)

\bibitem[]{}
Sembach, K. R., et al. 2003, ApJS, 146, 165

\bibitem[]{}
Spitzer, L., 1978, Physical Processes in the Interstellar Medium
(New York: Wiley).

\bibitem[]{}
Thomas, D., Maraston, C., \&  Bender, R. 2003, MNRAS, 343, 279 (T03)

\bibitem[Totani et al.(2008)]{Totani_08}
Totani, T., Morokuma, T., Oda, T., Doi, M., \& Yasuda, N. 2008, PASJ, 60, 1327

\bibitem[]{}
Tsujimoto, T. 2011, ApJ, 736, 113

\bibitem[]{}
Tsujimoto, T., \& Bekki, K. 2012, ApJ, 736, 113

\bibitem[]{}
Tsujimoto, T., Nomoto, K., Yoshii, Y., Hashimoto, M., Yanagida, S.,
\& Thielemann, F.-K. 1995, MNRAS, 277, 945 (T95)

\bibitem[]{}
Van der Swaelmen, M.,  Hill, V.,   Primas, F., \&  Cole, A. A.
2013, A\&A, 560, 44

\bibitem[]{}
Vazdekis, A., et al., 2010, MNRAS, 404, 1639

\bibitem[]{}
Welty, D. E., Hobbs, L. M., Lauroesch, J. T., Morton, D. C.,
Spitzer, L.,   York, D. G., 1999, ApJS, 124, 465

\bibitem[]{}
Winters, J. M., Fleischer, A. J.,  Le Bertre, T.,   Sedlmayr, E., 
1997, A\&A, 326, 305


\bibitem[]{}
Xilouris, E., Alton, P., Alikakos, J., Xilouris, K., Boumis, P.,  Goudis, C.,
2006, ApJ, 651, L107

\bibitem[]{}
Yoshida, M., Kawabata, K. S.,   Ohyama, Y., 2011, PASJ, 63, 493

\bibitem[]{}
Yozin, C.,  Bekki, K., 2012, ApJ, 756. L18

\bibitem[]{}
Yozin, C.,  Bekki, K., 2014, MNRAS, 443, 522

\end{thebibliography}
\end{document}